\documentclass[USenglish,nolipics]{lipics-v2018}

\usepackage{microtype}
\usepackage{amsmath}
\usepackage{graphicx}
\usepackage{indentfirst}
\usepackage{amsfonts,amssymb}
\usepackage{algorithm}
\usepackage{adjustbox}

\usepackage[noend]{algpseudocode}
\usepackage{accents}
\usepackage{color,soul}
\usepackage{mathtools}
\usepackage{enumitem}
\setlist[itemize]{itemsep=0.5em, topsep=0.5em, parsep=0pt, partopsep=0pt}
\graphicspath{ {./images/} }

\title{Beyond Worst-Case Subset Sum: An Adaptive, Structure-Aware Solver with Sub-$2^{n/2}$ Enumeration}

\author{Jesus Salas}{  }{jesus.salas@gmail}{https://orcid.org/0009-0007-6411-2270}{}{}
\authorrunning{An Adaptive and Structure-Aware Subset Sum Solver}
\titlerunning{An Adaptive and Structure-Aware Subset Sum Solver}

\subjclass{NP-complete problems, Design and analysis of algorithms}

\nolinenumbers 
\hideLIPIcs  

\begin{document}

\begin{center}
  {\large\bfseries Beyond Worst-Case Subset Sum: An Adaptive, Structure-Aware Solver with Sub-$2^{n/2}$ Enumeration}\\[1.5em]
  {\normalsize Jesus Salas}\\[0.5em]
  {\small Independent Researcher}\\[0.5em]
  {\small jesus.salas@gmail}\\[0.5em]
  {\small \url{https://orcid.org/0009-0007-6411-2270}}\\[2em]
\end{center}
\begin{abstract}
The Subset Sum problem, which asks whether a given set of $n$ integers contains a subset summing to a specified target $t$, is a fundamental NP-complete problem arising in cryptography, combinatorial optimization, and beyond. The classical meet-in-the-middle (MIM) algorithm of Horowitz and Sahni achieves a worst-case time complexity of $O^*(2^{n/2})$, which remains the best-known deterministic bound. Yet many instances exhibit abundant collisions in partial sums, so the true difficulty is often governed by $U$, the number of unique subset sums $U = |\Sigma(S)|$.

We present a structure-aware, adaptive solver that deterministically enumerates only the distinct sums, pruning duplicates on the fly. Its runtime is parameterized by $U$, the number of unique subset sums, achieving $O^*(U \cdot n^2)$ performance. Its core is a unique-subset-sums enumerator combined with a double meet-in-the-middle strategy and lightweight dynamic programming, thereby avoiding the classical MIM’s expensive merging phase.

Additionally, we introduce a Controlled Aliasing technique that ensures strictly sub-$2^{n/2}$ enumeration even on unstructured instances with near-maximal subset sum diversity, effectively reducing the exponent by a small but nontrivial constant factor. Our solver supports anytime and online operation, producing intermediate solutions (partial expansions) early and adapting seamlessly to newly added elements.

Both theoretical analysis and empirical evidence show that for structured inputs—such as those with small doubling constants, high additive energy, or significant additive redundancy—our method can significantly outperform classical approaches, often approaching near-dynamic programming efficiency. Even in the worst-case regime, it never exceeds the $O^*(2^{n/2})$ bound, and its controlled aliasing-based pruning guarantees a genuine constant-factor speedup over naive MIM enumerations.

\end{abstract}
\noindent
\section{Introduction}

\footnotetext{This paper presents the complete algorithmic framework and empirical validation of the solver. The theoretical implications of this approach—including its connection to Instance Complexity and fine-grained hardness—are explored in a companion paper.}

\smallskip

The Subset Sum problem is a classical NP-complete problem that asks whether a given set \(S\) of \(n\) integers contains a subset whose elements sum to a specified target \(t\). Traditionally, worst-case complexity is measured in terms of \(n\) and the total number of potential subsets, i.e., \(2^n\). However, in many practical scenarios, the input exhibits significant \textit{additive structure}—captured by parameters such as the \emph{doubling constant} or \emph{additive energy}—which implies that many subsets yield the same sum. In such cases, the true computational challenge is determined not by \(n\) alone but by the \emph{effective search space} \(U\), defined as the number of unique subset sums in \(\Sigma(S)\).

\smallskip
The seminal meet-in-the-middle algorithm of Horowitz and Sahni~\cite{horowitz-sahni-1974} achieves a worst-case running time of \(O^*(2^{n/2})\) by dividing \(S\) into two lists and enumerating all subset sums for each. Although this bound remains tight in theory, empirical studies and recent research~\cite{bringmann_nakos_2020,randolph2024parameterizedalgorithmsintegersets} suggest that the practical difficulty of an instance is better characterized by the number of unique subset sums rather than the total number of subsets. For example, when the input set has a low doubling constant or high additive energy, many subset sums coincide, reducing \(U\) and, consequently, the work required to solve the instance.

\smallskip
A significant body of work has analyzed the \emph{empirical hardness} of NP-complete problems, including Subset Sum and related variants such as k-Sum or Knapsack. Early investigations by \cite{cheeseman-kanefsky-taylor} illustrated how “phase transitions” can yield exceptionally hard instance families, even though many randomly generated instances are surprisingly easy in practice. Subsequent works \cite{lagarias-odlyzko, coster-et-al, flaxman-przydatek} have shown that the \emph{density} of Subset Sum instances (often measured by \(n / \log(\max a_i)\)) strongly influences typical-case runtimes. In particular, so-called “low-density” or “medium-density” inputs can often be solved faster than the classical \(2^{n/2}\) meet-in-the-middle bound would suggest. Recent studies in cryptographic settings \cite{howgrave-graham-joux-2010} further document large empirical speedups over the worst-case analysis, confirming that the actual difficulty of Subset Sum hinges dramatically on its structural properties and input distributions.

\smallskip
Randomization can also help in practice. Several contributions have aimed to improve the time complexity for solving Subset Sum. In particular, \textbf{randomized algorithms} introduced in~\cite{howgrave-graham-joux-2010, bringmann_nakos_2020} leverage the additive structure to reduce the effective search space. Although these randomized methods often yield practical speedups, they typically introduce an element of uncertainty and do not provide a deterministic guarantee of an exact solution.

\smallskip
These empirical findings reinforce our structure-sensitive perspective: they demonstrate that for many real or random-like instances (whether sparse or dense), the effective search space may be significantly smaller than \(2^n\).

\smallskip
Despite these advances, a universal method that deterministically exploits the combinatorial structure of Subset Sum across all instances remains elusive. The challenge arises from the diversity of inputs: on the one hand, \emph{collision-rich} (or small-range) instances can be tackled effectively by pseudopolynomial DP or meet-in-the-middle; on the other hand, \emph{sparse} sets with minimal collisions may require different techniques (e.g., sparse convolution). Moreover, although results such as Freiman’s Theorem provide strong bounds for sets with small doubling constants, they do not extend uniformly to all instances.

\smallskip
Motivated by these observations, we propose a structure-aware framework that leverages both the combinatorial properties of the enumeration process and the underlying additive combinatorics of the Subset Sum problem. Our approach centers on a novel \emph{unique-subset-sums enumerator} that generates each distinct subset sum exactly once, thereby parameterizing the runtime by the effective search space size \(U = |\Sigma(S)|\).

\smallskip
In addition, we introduce a \emph{Controlled Aliasing} mechanism that syntactically forces early collisions during subset enumeration. This guarantees a worst-case runtime that is \textbf{provably and strictly faster than the classical \(O^*(2^{n/2})\) bound}, even on unstructured instances with near-maximal subset sum diversity. These reductions are achieved without modifying the input or losing solution completeness. They are purely structural improvements to the enumeration process.

\smallskip
Our solver further supports \emph{anytime} and \emph{online} operation: it can output partial enumerations, resume computation efficiently, and incorporate new elements without reinitializing. This makes the approach suitable for real-time applications and interactive solvers.

\smallskip

Altogether, the solver presents a unified, output-sensitive framework that adapts gracefully to the structure of each input. It provides rigorous improvements over classical exponential-time algorithms, not only on structured instances but also provably on worst-case inputs. This paper is dedicated to the full algorithmic description and empirical validation of this solver. The broader theoretical implications of this structure-aware approach are explored in a series of companion papers— the IC-SubsetSum~\cite{ic-subsetsum}, which formalizes the certificate-sensitive framework underlying this solver, and its extensions to Knapsack~\cite{ic-knapsack} and 3-SAT~\cite{ic-3sat}.
\section{Key Contributions}

In this work, we present a novel adaptive and structure-aware solver for the Subset Sum problem that dynamically adjusts to the instance structure. The key contributions of our approach are:

\begin{itemize}
    \item \textbf{Unique Subset Sums Enumeration:} We introduce an enumeration strategy that generates each unique subset sum exactly once, eliminating redundant computations and dynamically pruning branches.

    \item \textbf{Double Meet-in-the-Middle Optimization:} Our solver leverages a double meet-in-the-middle approach to efficiently combine solutions online from two halves, avoiding expensive sorting and merging phases.

    \item \textbf{Controlled Aliasing:} We propose a lightweight structural optimization that forces collisions in the enumeration tree, guaranteeing a \textbf{worst-case runtime that is provably and strictly faster than the classical \(O^*(2^{n/2})\) bound}.

    \item \textbf{Anytime and Online Behavior:} The solver supports incremental computation and allows seamless updates when new elements are introduced, making it suitable for real-time applications.

    \item \textbf{Output-Sensitive and Adaptive Runtime:} The total runtime scales with the number \( U \) of unique subset sums, \textbf{achieving a performance that is polynomial in \(n\) and \(U\)}.

    \item \textbf{A Unified Framework for Adaptive Enumeration:} The solver's design provides a concrete and unified framework for exploiting instance-specific structure, serving as a blueprint for other adaptive, deterministic exponential-time algorithms.

\end{itemize}

\section{Preliminaries}
\label{sec:prelim}

In this section, we introduce the basic concepts, notations, and parameters that form the foundation for our analysis and algorithm.

\subsection{Basic Definitions}

Let \(S \subset \mathbb{Z}\) be a nonempty finite set of \(n\) integers. The classical \emph{Subset Sum Problem} is defined as follows:

\begin{definition}[Subset Sum Problem]
Given a set \(S\) of \(n\) integers and a target \(t\) satisfying
\[
1 \le t \le \sum_{x \in S} x,
\]
determine whether there exists a subset \(S' \subseteq S\) such that
\[
\sum_{x \in S'} x = t.
\]
\end{definition}

\begin{definition}[Asymptotic Notation]
We use the notation $O^*(f(n))$ to denote an upper bound that suppresses polynomial factors.
\end{definition}

\begin{definition}[Density]
We define the \emph{density} of a Subset Sum instance \(S = \{a_1, a_2, \dots, a_n\}\) as
\[
d(S) = \frac{n}{\log_2 (\max S)},
\]
which measures the ratio between the number of elements and the bit-length of the largest element. A higher density indicates that many elements are packed into a relatively small numerical range, often leading to more additive collisions and a smaller effective search space.
\end{definition}

\begin{definition}[\(k\)-Permutations and \(k\)-Subsets]
For a set \(S\), a \emph{\(k\)-permutation} is an ordered selection of \(k\) distinct elements from \(S\) representing a partial solution in the enumeration process. When order is not significant, the same collection is referred to as a \emph{\(k\)-subset}. In our algorithm, we generate \(k\)-permutations incrementally, and upon verifying that their aggregated sum is unique, we interpret them as \(k\)-subsets.
\end{definition}

\begin{definition}[Enumeration Tree Topology]
The \emph{enumeration tree topology} of an instance \(S\) refers to the structure of the combinatorial tree formed by all possible subsets of \(S\). This topology is determined by the additive relationships within \(S\), and remains fixed for a given static instance. It dictates how distinct \(k\)-permutations may converge to the same subset sum.
\end{definition}

\subsection{Unique Subset Sums}

Although there are \(2^n\) possible subsets of \(S\), many of these subsets yield the same sum when the input set \(S\) exhibits additive structure. We define the set of all \emph{unique subset sums} as:
\[
\Sigma(S) = \left\{ \sum_{x \in S'} x \,:\, S' \subseteq S \right\}.
\]
The \emph{effective search space} is then given by:
\[
U = |\Sigma(S)|.
\]
In many practical instances, \(U\) is significantly smaller than \(2^{n/2}\), and our algorithm explicitly exploits this fact by parameterizing the running time in terms of \(U\).

\smallskip
\noindent
Note that although \(|\Sigma(S)| \le 2^n\) in general, our solver enumerates only one side of the meet-in-the-middle split at a time. This naturally limits the number of generated sums to \( U \le 2^{n/2} \), which serves as the relevant upper bound throughout our analysis.

\begin{remark} For clarity, we assume throughout that $S \subset \mathbb{Z}_{\ge 0}$. The methods generalize naturally to inputs containing $0$ (introducing a trivial collision) or negative values (handled by symmetric memoization). In the worst case, storing all unique subset sums requires $O^*(2^{n/2})$ space, though this cost is typically much lower in collision-rich instances.
\end{remark}

\subsection{Additive Structure Measures}

The following measures quantify the degree of additive structure in the set \(S\):

\paragraph*{Doubling Constant}
The \emph{doubling constant} of \(S\) is defined by
\[
C = \frac{|S+S|}{|S|}, \quad \text{where } S+S = \{a+b \mid a,b \in S\}.
\]
A low doubling constant indicates that many pairwise sums overlap, which in turn suggests that the effective number \(U\) of unique subset sums is reduced.

\paragraph*{Additive Energy}
The \emph{additive energy} \(E(S)\) is a measure of the number of additive collisions in \(S\). It is defined as the number of solutions in \(S\) to
\[
a + b = c + d, \quad \text{with } a,b,c,d \in S.
\]
High additive energy implies that many subsets yield the same sum, again reflecting a small \(U\).

\subsection{Extended Structural Parameters}

Beyond the doubling constant \(C\) and the additive energy \(E(S)\), we will also make use of three additional parameters to quantify the structure or redundancy in \(S\).

\paragraph*{Linearity Factor \(\lambda\).}
We define a \emph{linearity factor} \(\lambda \in [0,1]\) to capture the extent to which the elements of \(S\) exhibit near-linear (e.g., arithmetic-progression-like) relationships. A smaller \(\lambda\) indicates that the set resembles a linear or nearly linear structure, often leading to higher collision rates among subset sums.

\paragraph*{Clustering Factor \(\gamma\).}
We define a \emph{clustering factor} \(\gamma \in [0,1]\) to indicate how tightly the distinct sums in \(\Sigma(S)\) cluster. Smaller \(\gamma\) implies that many subset sums overlap or collide numerically, thus reducing the effective search space \(U\).

\paragraph*{Duplicate Measure \(\delta\).}
We let \(\delta\) represent the effective number of duplicate elements in \(S\). Repeated elements often cause multiple branches in our enumerator to converge onto the same partial sum, effectively \emph{merging} those branches rather than keeping them distinct. Consequently, having more duplicates reduces the effective search space \(U\).

\smallskip
\noindent
These three parameters (\(\lambda\), \(\gamma\), and \(\delta\)) all capture different aspects of \emph{additive redundancy} in \(S\). In particular, they complement the more classical measures \(C\) and \(E(S)\) by providing additional insight into how and why the number of unique subset sums \(U\) might remain well below \(2^n\).

\paragraph*{Remark on ``Input-Specific'' Hardness.}
Throughout this paper, when we discuss ``input-specific'' or ``structure-aware'' complexity, we are referring to significant \emph{subfamilies} of instances that exhibit certain structural properties (e.g., high collision rates, small doubling constants, or additive redundancy). These properties can cause the number of unique subset sums, \(U = |\Sigma(S)|\), to remain far below the naive \(2^n\), making the instance practically easier than worst-case bounds would suggest. This perspective does not contradict NP-completeness; it simply underscores that many real or structured inputs deviate markedly from the fully unstructured worst-case distribution, thus giving the solver room for substantial pruning.

\subsection{Additive Structure and the Effective Search Space}

The performance of our solver is governed by the parameter $U = |\Sigma(S)|$, the number of unique subset sums induced by the input set $S$. While the worst-case value of $U$ is $2^n$, most practical instances exhibit substantial additive redundancy that causes $U \ll 2^n$, reducing the effective search space and accelerating the solver.

This redundancy arises from various structural properties of $S$, which can be quantified by both classical and extended measures:

\begin{itemize}
    \item \textbf{Doubling Constant $C$}: A small doubling constant $C = |S + S| / |S|$ implies that many pairwise sums overlap, compressing the sumset $\Sigma(S)$.
    \item \textbf{Additive Energy $E(S)$}: A high number of additive collisions $a + b = c + d$ indicates that many subsets share the same sum.
    \item \textbf{Linearity Factor $\lambda$}: When elements of $S$ are nearly linear (e.g., forming arithmetic progressions), many subset sums coincide.
    \item \textbf{Clustering Factor $\gamma$}: Tightly bunched subset sums reflect a high degree of numerical collision.
    \item \textbf{Duplicate Measure $\delta$}: Repeated elements merge branches of the enumeration tree, reducing the number of distinct paths and hence unique sums.
\end{itemize}

Each of these parameters contributes to a smaller $U$, and together they provide a robust explanation for why many structured instances are significantly easier than the $2^n$ worst case would suggest.

The solver exploits this structure explicitly by generating each unique subset sum exactly once and pruning further expansions when collisions are detected. In unstructured instances, where $U$ approaches $2^{n/2}$, the solver degrades gracefully to the classical meet-in-the-middle performance.

We do not develop these tools formally in this work. For rigorous treatments, we refer the reader to foundational texts in additive combinatorics such as \cite{tao2006additive, tao2010structure, ruzsa1994sumsets, green2007freiman} and to the companion theory papers that connect the algorithm runtime directly to these measures.

\section{Prior Work}
\label{sec:prior}

The Subset Sum problem has attracted extensive research due to its NP-completeness and practical significance. Early approaches, such as Bellman’s dynamic programming algorithm~\cite{bellman-1954}, offered pseudopolynomial-time solutions with time complexity \(O(n\,t)\) (where \(t\) is the target sum). Over time, several refined DP-based techniques have been proposed to mitigate this cost when \(t\) remains large. For instance, Koiliaris and Xu~\cite{koiliaris-xu-2017} introduced a faster pseudopolynomial-time algorithm running in \(\widetilde{O}(t\sqrt{n})\) by leveraging FFT-based convolutions for partial-sum computations. More recently, Bringmann~\cite{bringmann2017near} developed a near-linear pseudopolynomial algorithm for Subset Sum under specific parameter regimes, achieving further speedups in cases where the numeric range is not excessively large. Nevertheless, once \(t\) grows beyond a moderate threshold, these DP-based methods become infeasible in practice because their complexity remains tied to the numerical size of the target.

\smallskip

A major breakthrough was achieved with the meet-in-the-middle algorithm of Horowitz and Sahni~\cite{horowitz-sahni-1974}, which achieves a worst-case time complexity of \(O^*(2^{n/2})\) by splitting the input set into two parts and enumerating all subset sums for each half. However, while this bound remains tight for pathological inputs, empirical evidence and more recent theoretical insights~\cite{bringmann2017near, nederlof2021improving} reveal that \emph{practical} Subset Sum hardness is more accurately governed by \emph{collisions} in partial sums rather than the full \(2^n\) enumeration.

\paragraph*{Exploiting Additive Structure}
A key observation in recent work is that many inputs exhibit significant \textit{additive structure}, which naturally leads to a high degree of such collisions. This structure can be quantified by measures such as the \emph{doubling constant} and \emph{additive energy}. For instance, Freiman's Theorem~\cite{freiman1964theorem} guarantees that sets with small doubling can be embedded in a low-dimensional generalized arithmetic progression (GAP), implying that the effective search space \(U\) is dramatically smaller than \(2^n\) when the input is highly structured. These observations motivate the use of adaptive runtime bounds based on \( U = |\Sigma(S)| \), as formalized in our deterministic \( O(U \cdot n^2) \) solver.

\paragraph*{Randomized Algorithms}
Randomized approaches~\cite{howgrave-graham-joux-2010, bringmann_nakos_2020} have been proposed to exploit additive structure by using collision-based pruning techniques. These algorithms navigate the combinatorial tree of subset sums probabilistically, effectively reducing the number of distinct paths that must be examined. 

\smallskip

A recent randomized algorithm by Bringmann, Fischer, and Nakos~\cite{bringmann2024subsetsumunconditional} achieves the first unconditional improvement over Bellman’s dynamic programming algorithm for Subset Sum, solving the target-constrained decision problem in $\widetilde{O}(|S(X, t)| \cdot \sqrt{n})$ time. Their approach focuses on computing reachable sums up to a target $t$, using algebraic and combinatorial techniques. 

\smallskip

While these methods often yield substantial practical speedups, they are often probabilistic (e.g., Monte Carlo) and thus provide a correct answer with high probability rather than with deterministic certainty.

\paragraph*{Challenges and the Need for Determinism}
Despite the improvements provided by randomized techniques, these methods generally do not extend well to all instances, especially sparse inputs where collisions are infrequent. Moreover, existing deterministic methods (such as the classical meet-in-the-middle approach) do not fully exploit the underlying additive structure of the input. There remains a gap in the literature for a deterministic method that adapts to additive structure while maintaining provable worst-case guarantees.

\paragraph*{Fractal Geometry and Structure-Aware Hardness}

Horn, van den Berg, and Adriaans~\cite{horn2024fractal} recently proposed a fractal analysis of Subset Sum, showing that the histogram of subset sums exhibits self-similarity, with fractal dimension correlating to practical hardness in branch-and-bound solvers. Their approach provides a geometric perspective on instance difficulty.

This view is complementary to ours: we focus on $U = |\Sigma(S)|$ as the effective search space, using additive-combinatorial structure to explain why $U \ll 2^n$ in compressible instances. While our method emphasizes collision-driven pruning, their fractal dimension captures global redundancy in solution space.

\paragraph*{Relation to Instance Complexity}
\label{subsec:instance-complexity}
An important theoretical angle for analyzing how NP-complete problems can be significantly easier on ``structured'' inputs is the notion of \emph{instance complexity} studied by Orponen, Ko, Sch\"oning, and Watanabe~\cite{OrponenKoSchoningWatanabe1994}. Informally, instance complexity measures the size of a ``special-case program'' that decides whether a particular input $z$ belongs to the language $A$, within a given time bound, while never misclassifying any other inputs (it may answer ``don't know'' on them). They prove that although NP-hard sets still have infinitely many ``intrinsically hard'' instances, many classes of instances admit specialized programs of lower complexity. This perspective suggests a formal basis for the empirical observation that many ``structured'' instances are easier than worst-case analysis would predict.

\paragraph*{Our Contribution}
Recent research~\cite{bringmann2017near, bringmann_nakos_2020} has indicated that parameterizing Subset Sum in terms of \(U\) (or related structural measures) can yield exponential speedups in practical cases. However, prior work has largely relied on randomized methods or specific assumptions on the input structure.

Our work fills this gap by introducing a \textbf{deterministic} structure-aware framework with deterministic runtime \( O(U \cdot n^2) \) that leverages the effective search space \(U\) as a parameter, thereby providing a unified approach that adapts seamlessly to both dense and sparse instances.

The collision-driven pruning in our unique subset-sums enumerator does more than simply reduce enumeration overhead; it also offers a real-time lens into the \emph{structure} of the input. Specifically, the enumerator’s memoization table tracks how often newly formed partial sums coincide with previously encountered sums. When collisions appear frequently and early, this strongly indicates that the input set \(S\) exhibits nontrivial additive or redundant structure, for example:
\begin{itemize}
  \item A small doubling constant (\( |S+S| \le C|S| \)),
  \item High additive energy (\(a+b = c+d\) collisions),
  \item Clustered or near-linear arrangements of elements,
  \item Repetitions or duplicates that force multiple branches onto the same sums.
\end{itemize}

Without any additional preprocessing, the enumerator naturally \emph{adapts} to collisions as they occur, effectively exploiting the input’s underlying structure (whether dense, sparse, or in-between) and automatically adjusting its work to reflect the true hardness \(U = |\Sigma(S)|\).

\section{Unique Subset Sums Enumerator}
\label{sec:enum}

A naive solution to the Subset Sum problem would exhaustively enumerate all \(2^n\) subsets of \(S\) and check whether any of them sum to the target \(t\). In practice, however, when the input set \(S\) exhibits significant \emph{additive structure}, many of these \(2^n\) subsets produce redundant sums. The true challenge lies in generating the \emph{unique subset sums}---that is, the effective search space of distinct sums defined by
\[
\Sigma(S) = \left\{ \sum_{x \in S'} x : S' \subseteq S \right\}.
\]
We denote the size of this set by 
\[
U = |\Sigma(S)|.
\]
In many instances, \(U\) is dramatically smaller than \(2^n\). The total runtime of our solver is therefore determined not by the total number of subsets (\(2^n\)), but by the number \(U\) of unique subset sums.

\smallskip
Our novel enumeration model provides access to all generated data (i.e., the subsets of \(S\) and their sums) throughout the process.

\smallskip
To implement the enumerator, we designed a novel permutation-generating algorithm. Conventional permutation generators output one full permutation per step (for example, by swapping two elements from the previous permutation). In contrast, our approach generates all possible permutations of \(S\) one column at a time (i.e., left-to-right). Here, a ``column" refers to the position within the \(k\)-permutation being expanded---analogous to depth in a tree or length in a prefix.

At each step, the algorithm emits the \emph{\(k\)-permutations} (or \(k\)-subsets) of increasing cardinality according to the following process:

\begin{center}
\textbf{Initial step:} Start with an \textbf{INPUT} list containing the empty \(k\)-permutation, \(\{\}\).\\[1mm]
\textbf{For each step:}\\[1mm]
\quad Extend each entry in the \textbf{INPUT} by one column by appending every element from \(S\) not already included, and add these expanded \(k\)-permutations to the \textbf{OUTPUT}.\\[1mm]
\quad Once all entries in \textbf{INPUT} have been processed, set \(\textbf{INPUT} \gets \textbf{OUTPUT}\) and repeat until no new \(k\)-permutations are generated.
\end{center}

For example, consider the set \(S = \{1,2,3,4,5\}\):

\begin{center}
\textbf{Column Expansion Sub-process 1}\\[1mm]
\(\textbf{Input} = \{ \{\} \} \quad \Longrightarrow \quad \textbf{Output} = \{\{1\}, \{2\}, \{3\}, \{4\}, \{5\}\}\)\\[2mm]
\(\textbf{Input} \gets \textbf{Output}\)\\[2mm]
\begin{large} --------- \end{large}\\[2mm]
\textbf{Column Expansion Sub-process 2}\\[1mm]
\(\textbf{Input} = \{\{1\}, \{2\}, \{3\}, \{4\}, \{5\}\} \quad \Longrightarrow \quad \textbf{Output} = \{\{1,2\}, \{1,3\}, \{1,4\}, \{1,5\}, \{2,1\}, \{2,3\}, \ldots\}\)
\end{center}

Since permutations in isolation do not directly solve the Subset Sum problem, our algorithm tracks the sum of each expanded \(k\)-permutation during the column expansion process. Before adding an expanded \(k\)-permutation to the \textbf{OUTPUT}, the algorithm checks whether its sum has already been encountered in the memoization table. If so, the algorithm compares the new subset's \emph{canonical representation} (typically its index-based bitmask) against the stored one. If the new subset is lexicographically smaller, it replaces the stored representative; otherwise, it is pruned. This ensures that for each sum, exactly one canonical subset is retained---preventing over-pruning in collision-rich instances and guaranteeing correctness.

Once this process completes, all possible \(k\)-permutations for \(k \le \frac{n}{2}\) are generated. We restrict the enumeration to \(k \le \frac{n}{2}\) because any subset with \(k > \frac{n}{2}\) elements can be obtained by computing its complement with respect to \(S\).

This strategy eliminates redundant enumeration and significantly reduces both computational overhead and memory usage, while ensuring that all unique subset sums---i.e., all elements of \(\Sigma(S)\) (and hence the effective parameter \(U\))---are captured.

For the complete pseudocode of our unique subset sums enumerator algorithm, please refer to Algorithm 1 in  Appendix~\ref{app:algorithms}.

\smallskip
\subsection{\(k\)-Permutations and \(k\)-Subsets and the Combinatorial Tree Structure}

In this paper, we use the terms \emph{\(k\)-permutation} and \emph{\(k\)-subset} interchangeably to refer to the same entity, with the interpretation depending on the phase of the algorithm:
\begin{itemize}
    \item When an entity appears in the \textbf{INPUT} or \textbf{OUTPUT} lists during any column expansion sub-process, it is treated as a \(k\)-subset. In this context, it represents a subset of \(k\) elements that sums to a unique value and has passed the Unique Subset Sum constraint, which is the primary focus of our solver.
    \item When the entity serves as a prefix for generating \((k+1)\)-permutations during the column expansion, it is viewed as a \(k\)-permutation.
    \item Just before validation against the Unique Subset Sum constraint, the expanded \((k+1)\)-permutation is reinterpreted as a \(k\)-subset; if it satisfies the constraint, it is then added to the \textbf{OUTPUT}.
\end{itemize}

This duality clarifies the logic behind both the enumerator and the solver. \(k\)-permutations allow us to efficiently traverse and query the combinatorial tree structure during candidate generation, while \(k\)-subsets provide the precise representations needed to verify unique subset sums. This dual treatment is central to our approach, ensuring that the enumeration process remains both dynamic and efficient.

\subsection{Seeding support}
An important feature of our enumerator is its support for \emph{seeding}. If the memoization table is pre-populated with unique subset sums (i.e., elements of \(\Sigma(S)\)) and their corresponding \(k\)-subsets, the enumerator can seamlessly resume processing from that point. This capability allows intermediate results from previous cycles to be preserved and re-utilized, thereby reducing redundant computation and enabling an anytime (or incremental) behavior.
\subsection{Convolution-Like Behavior in \(k\)-Permutation Enumeration}

Although our unique subset sum enumerator does not perform convolution in the strict numerical sense (e.g., via the FFT), it exhibits behavior that is conceptually analogous to convolution. Traditional convolution combines two sequences \(A\) and \(B\) to produce a new sequence:
\[
A+B = \{a+b \mid a \in A,\ b \in B\}.
\]
In our approach, the enumeration process generates \(k\)-permutations, which represent all possible orderings of \(k\)-subsets of the input set \(S\). This process has several key characteristics that resemble convolution:

\begin{itemize}
    \item \textbf{Aggregation of Partial Sums:} As the algorithm extends partial \(k\)-permutations by appending new elements, it aggregates the resulting partial sums in a manner analogous to convolution summing overlapping portions of signals. This aggregation naturally leads to the identification of collisions—i.e., multiple \(k\)-permutations yielding the same subset sum.
    
    \item \textbf{Propagation Through the Combinatorial Tree:} Different orderings of the elements propagate the partial sums through various branches of the combinatorial tree. Importantly, many distinct paths converge on the same \(k\)-subset sum, since the order of elements does not affect the numerical total. This convergence mirrors how convolution combines contributions from shifted versions of a sequence, ensuring that all possible paths leading to the same sum are effectively merged.
    
    \item \textbf{Implicit Combination of Branches:} The self-similarity and structural properties of \(k\)-permutations ensure that once a particular subset sum is computed, further redundant branches that would yield the same sum are pruned. Much like convolution naturally consolidates overlapping components, our method consolidates identical subset sums, thereby reducing the effective search space.
\end{itemize}

Thus, while our algorithm does not explicitly implement FFT-based numerical convolution, it performs a topological convolution whose effects are expressed numerically: as the $k$-permutation enumeration process sweeps through the combinatorial space, it aggregates partial solutions, causing many distinct orderings to converge to the same unique subset sum.

\subsection{A 4-Column Litmus Test for Instance Hardness and Splitting}
\label{sec:litmus}

A practical method to estimate how ``collision-rich'' a Subset Sum instance is---and hence how small the effective search space $U = |\Sigma(S)|$ might be---is to run our enumerator for a bounded number of columns. In particular, limiting enumeration to subsets of size at most 4 yields a quick snapshot of the instance’s additive redundancy.

\paragraph*{Local collision proxy.} By enumerating all $k$-subsets of size at most 4 from either half of the input (of size $n/2$), we can measure how often partial sums collide. Frequent early collisions suggest that the full instance will have a significantly compressed sumset $\Sigma(S)$.

\paragraph*{Guidance for splitting.} This 4-column test can be used symmetrically on the two halves $\ell_0$ and $\ell_1$ to approximate the distribution of unique sums in each. If one side is significantly more collision-rich than the other, elements can be reassigned across the split to balance the enumeration cost and improve pruning in the main solver.

\paragraph*{Global application.} Alternatively, the test can be applied to the entire set $S$ to assess global structure before partitioning. This is especially useful when evaluating input hardness without executing the full solver.

A formal analysis of this approximation test and its runtime cost appears in Section~\ref{sec:litmus-runtime}.

\subsection{Correctness and Completeness of Enumerating Unique Sums}

\begin{theorem}[Enumerator Correctness]
\label{thm:enumerator-correctness}
Let \(\mathcal{E}\) be the enumerator described above, which expands partial solutions one element at a time, and resolves duplicate sums by storing the lexicographically minimal canonical subset representation. Then:
\begin{enumerate}
\item \textbf{(Completeness)} For every subset \(S' \subseteq S\), its sum \(\sum_{x \in S'} x\) eventually appears in the enumerator's memoization table.
\item \textbf{(Uniqueness)} Each distinct sum in \(\Sigma(S)\) is associated with exactly one canonical subset.
\end{enumerate}
\end{theorem}

\begin{proof}
We prove each property separately.

\paragraph*{(1) Completeness.}
Let \(S' = \{x_{i_1}, x_{i_2}, \dots, x_{i_k}\}\) be any subset of \(S\), written
such that \(i_1 < i_2 < \cdots < i_k\). We argue by induction on \(k\) that the enumerator
generates the partial sum \(\sum_{x \in S'} x\) at some stage.

\smallskip
\textit{Base Case} (\(k=0\)): The empty subset has sum 0, which is included during initialization.

\textit{Inductive Step}: Suppose all subsets of size at most \(k\) are correctly generated. For a subset \(S'\) of size \(k+1\), remove the last element to obtain \(S''\). By the hypothesis, \(\sum S''\) appears in the memo. During expansion, \(x_{i_{k+1}}\) is eventually added, producing \(\sum S'\). If it’s new, it is inserted; if it collides, the canonical comparison ensures the best representative is retained.

\paragraph*{(2) Uniqueness.}
Assume for contradiction that the memoization table stores the same sum \(s\) with two different subsets \(A \neq B\). When \(B\) is expanded, the algorithm checks \(s\) in the memo, finds it stored by \(A\), and compares canonical representations. Since the table always keeps only the lex-minimal one, either \(B\) replaces \(A\), or it is discarded. Thus, only one canonical subset is ever stored per sum.
\end{proof}

\subsection{Complement Trick for Half-Enumeration}
\label{subsec:half-enum-complement}

In our solver, the enumerator in each split is restricted to generate only subsets of size at most $\lfloor n/4 \rfloor$, where the split size is $n/2$. This bounded expansion supports top-down enumeration, reduces memory pressure, and simplifies implementation. The following lemma shows that this restriction does not compromise correctness.

\begin{lemma}[Complement Trick for Half-Enumeration]
\label{lem:bounded-subset-size}
Let $\ell$ be one half of the input, containing $n/2$ elements. Suppose the enumerator enumerates only subsets of size $\leq \lfloor n/4 \rfloor$. Let $U_{\max}$ be the number of unique subset sums discovered. Then for any subset $S' \subseteq \ell$ of size $> n/4$, its sum can be reconstructed via complement logic using previously enumerated subsets of size $\le n/4$.
\end{lemma}

\begin{proof}[Proof Sketch]
Let $S' \subseteq \ell$ be a subset of size $> n/4$. Then its complement $S^* = \ell \setminus S'$ has size $< n/4$, since $\ell$ has size $n/2$. The enumerator processes all such small subsets and stores their sums. Let $\sigma^* = \sum_{x \in S^*} x$ be the recorded sum. Then:

\[
\sum_{x \in S'} x = \sum_{x \in \ell} x - \sigma^*
\]

Since the total sum $\sum_{x \in \ell} x$ is precomputed once, every sum arising from a large subset $S'$ is recoverable via complement subtraction. The canonical bitmask for $S'$ is inferred as the complement of the stored representative for $S^*$.

See Section~\ref{sec:solver} for how complements are integrated during solution checking.
\end{proof}

\subsection{Implementation Invariants}

A correct and efficient implementation maintains the following key invariants:

\begin{itemize}
  \item A queue (or list) of partial $k$-permutations for current expansion.
  \item A memoization hash table $\text{Memo}$ mapping each unique subset sum $\sigma$ to:
  \begin{itemize}
    \item (i) a flag (or usage counter), and
    \item (ii) a pointer or bitmask representing the canonical $k$-subset that generated $\sigma$.
  \end{itemize}
  \item For each $\sigma$ in $\text{Memo}$:
  \[
  \sigma = \sum_{x \in X} x \quad \text{for some unique, canonical } X \subseteq S.
  \]
  \item Before expanding any $k$-permutation by adding $x$, the algorithm checks whether $\sigma + x$ is already in $\text{Memo}$. If so, it is pruned or replaced only if the new subset is lex-smaller.
\end{itemize}

These invariants guarantee that each partial sum is processed at most once and that the memoization table contains no redundant or superseded encodings. Hash-table operations are assumed to be $O(1)$ amortized.

\paragraph*{Collision Detection.} Once $\sigma$ is inserted, further attempts to create $\sigma$ via other permutations are rejected or skipped. This pruning ensures that work is proportional to $U$, not $2^n$.

\subsection*{Relation to BFS-Based Enumeration}

Although our method may superficially resemble BFS in that it proceeds incrementally, its structure and semantics differ fundamentally.

\begin{itemize}
  \item \textbf{Column-by-Column vs.\ Layer-by-Layer:}  
    BFS typically expands all subsets of size $k$ before moving to $k+1$. In contrast, our enumerator expands $k$-permutations one column at a time and prunes duplicates immediately upon detecting a sum already in the memo table.

  \item \textbf{Adaptive Scheduling:}  
    Our column-wise expansion is governed by a dynamic look-ahead policy (see Section~\ref{sec:divide_and_conquer}), which allows subsets to be deferred across cycles. Traditional BFS has no such per-permutation control or incremental delay.

  \item \textbf{Numerical and Structural Convolution:}  
    Our method incorporates real-time structural pruning based on partial sums, not just subset cardinality. This induces early merging in the combinatorial tree, collapsing branches that BFS would still explicitly explore within a layer.
\end{itemize}

While some BFS-style methods (e.g., Schroeppel–Shamir) introduce pruning or memory savings, they still operate on fixed layers and do not naturally support anytime scheduling, dynamic subset prioritization, or on-the-fly canonical selection. Thus, our approach should be viewed not as a refinement of BFS but as a structurally different enumeration regime rooted in memoized collision detection and partial-sum uniqueness.

\section{Subset-Sum Solver with Double Meet-in-the-Middle Optimization}
\label{sec:solver}

Having described our foundational Unique Subset Sums enumerator—which generates \(k\)-subsets (for \(k \le \frac{n}{2}\) per split) and produces only the unique subset sums (of effective cardinality \(U = |\Sigma(S)|\))—we now integrate it into a complete Subset-Sum solver.

\smallskip
We begin by splitting the input set \(S\) into two disjoint subsets, \(\ell_0\) and \(\ell_1\), each of size \(n/2\). For each split \(\ell_x\) (where \(x \in \{0,1\}\)), we run our enumerator to generate all unique subset sums, reducing the effective search space from \(2^{n/2}\) potential sums in each half down to \(U_0 = |\Sigma(\ell_0)|\) and \(U_1 = |\Sigma(\ell_1)|\). In many practical instances, both are substantially smaller than the worst-case bound.

\smallskip
We do \emph{not} need to wait until each enumerator completes in order to check for a viable solution. Instead, as \emph{soon} as a new unique subset sum is produced in one split, we verify in real time whether it completes a solution. This includes checking for direct matches, complements, and merges with sums from the other half—what we term a \emph{double meet-in-the-middle} strategy. On average, verifying each new partial sum requires a constant-time lookup (amortized \(O(1)\) with hashing) to determine whether it can combine with a counterpart to yield the target.

\smallskip
This on-the-fly procedure is supported by the following completeness lemmas:

\smallskip
Let \(\sigma(A) := \sum_{a \in A} a\) denote the sum of elements in a subset \(A \subseteq S\), in contrast to \(\Sigma(A)\), which refers to the full sumset of all subsets of \(A\)

\medskip
\noindent\textbf{Lemma 1.}  
If a subset \( A \subseteq \ell_x \) satisfies \( \sigma(A) = t \), then a valid solution is found without involving the other split.

\medskip
\noindent\textbf{Lemma 2.}  
Let \(A \subseteq \ell_x\) and \(B \subseteq \ell_y\), where \(\ell_0 \cup \ell_1 = S\). Then
\[
\sigma(A \cup B) = t \quad \Longleftrightarrow \quad \sigma(B) = t - \sigma(A).
\]

\medskip
\noindent\textbf{Lemma 3.}  
For each split \(\ell_x\) independently, the complement subset \(A^c = \ell_x \setminus A\) yields a valid solution if
\[
\sigma(A^c) = \sigma(\ell_x) - t.
\]

\medskip
\noindent\textbf{Lemma 4.}  
If the complements of subsets from both splits (denoted \(A^c \subseteq \ell_0\) and \(B^c \subseteq \ell_1\)) satisfy
\[
\sigma(A^c \cup B^c) = \sigma(S) - t,
\]
then a valid solution is found.

\medskip
\noindent\textbf{Lemma 5.}  
If a subset \(A \subseteq \ell_0\) and the complement of a subset \(B^c \subseteq \ell_1\) satisfy
\[
\sigma(A \cup B^c) = t,
\]
then a valid solution is obtained.

\medskip
These lemmas cover all structural configurations in which a subset of \(S\) can sum to \(t\), and are applied dynamically during enumeration. By working with the unique subset sums (of effective size \(U\)) rather than all \(2^{n/2}\) candidates, the number of combinations considered is dramatically reduced, especially in collision-rich instances.

\smallskip
By merging complementary sums from each half, our method verifies candidate solutions in constant time on average, thereby avoiding the sort/merge step of classical meet-in-the-middle approaches.

\smallskip
For the complete pseudocode of our double meet-in-the-middle algorithm, please refer to Algorithm~2 in Appendix~\ref{app:algorithms}.

\subsection{Time Complexity Analysis}
\label{sec:time-complexity}

The enumeration algorithm maintains a dynamic frontier of reachable sums across both halves of the input. Each side’s unique sums are generated via the structure-aware subset sum enumerator described earlier, and combinations are performed only on compatible pairs. For every sum \(s\) from the left half, the algorithm checks for a corresponding sum \(t - s\) in the right half, or other lemma-induced matches.

\smallskip
In the worst case, this results in roughly \(\binom{n/2}{n/4}\) partial sums per side, and a quadratic number of combinations---yielding the familiar \(O^*(2^{n/2})\) bound. However, the key feature of our approach is that we avoid generating duplicate sums at every step. Each intermediate expansion is filtered through a hash set or sorted list to ensure uniqueness, and combination occurs only over distinct elements.

\smallskip
Thus, the overall time complexity of the solver is \(O^*\left( \binom{n/2}{n/4} \right) \approx O^*(2^{n/2})\). In structured instances where the number of unique subset sums is small, the runtime is governed by enumerating these sums. Let \(U_0 = |\Sigma(\ell_0)|\) and \(U_1 = |\Sigma(\ell_1)|\) denote the number of distinct subset sums in each half. Then the total runtime is \(O^*(\max\{U_0, U_1\} \cdot n^2)\) deterministically, or \(O^*(\max\{U_0, U_1\} \cdot n)\) in expectation with randomized hashing. For clarity, we often write this as \(O^*(U \cdot n^2)\), where \(U = |\Sigma(S)|\), but the algorithm operates without ever constructing \(\Sigma(S)\) explicitly.

\smallskip
This cost includes both the subset expansions and the canonical comparison steps required to maintain correctness, which contribute to the runtime in collision-rich instances.

\paragraph*{Formal Analysis.} A complete proof of the solver’s runtime---including the derivation of the \(O(U \cdot n^2)\) deterministic bound and the \(O(U \cdot n)\) expected randomized bound---is presented in the companion theory paper~\cite{ic-subsetsum}. There, we analyze the memoization structure, collision pruning, and controlled expansion logic in detail under the certificate-sensitive model.

\subsubsection*{Runtime of Structure Detection via 4-Column Test}
\label{sec:litmus-runtime}

The 4-column test introduced in Section~\ref{sec:litmus} can be implemented efficiently. When applied to a half of size \(n/2\), it enumerates
\[
\sum_{k=0}^4 \binom{n/2}{k} = O(n^4)
\]
total subsets. Each subset sum is computed and recorded with a canonical representation (e.g., index tuple of size \(\leq 4\)). Since both sum computation and canonical comparison are bounded by \(O(1)\) per subset, the total runtime is:

\begin{itemize}
  \item \textbf{Deterministic runtime:} \(O(n^4)\), due to bounded comparisons of size-\(\leq 4\) subsets.
  \item \textbf{Randomized expected runtime:} \(O(n^4)\), using hash-based memoization.
\end{itemize}

This approximation can serve as a preprocessing pass to estimate instance hardness before full enumeration, or to inform input partitioning strategies.

\section{Meet-in-the-Middle Speed-Up via Controlled Aliasing}
\label{sec:duplicate-speedup}

Meet-in-the-middle (MIM) is a classical approach to \textsc{Subset~Sum} (and related
problems) that splits the input set $S$ of size $n$ into two lists of size $\approx n/2$ each.
It then enumerates all $2^{n/2}$ possible subsets in each half, producing two lists of
partial sums, and merges them in time $O^*(2^{n/2})$
\cite{horowitz-sahni-1974}. Below, we describe a simple yet effective technique that reduces the exponent by a noticeable
\emph{constant} factor, while preserving correctness. 

\subsection*{Controlled Aliasing}

We can shrink the exponent by a constant factor by syntactically forcing a collision within the enumeration logic of each split. This does not alter the original problem set \( S \); rather, it reshapes the enumeration tree to prune redundant paths earlier.

Consider two distinct elements in one half of the split, say \( x_0 \) and \( x_1 \). During the enumeration of subset sums for this half, we can apply a temporary rule: whenever a subset includes \( x_1 \), we compute its contribution to the sum as if it were \( x_0 \).

This aliasing means that the distinct subsets \( \{x_0\} \) and \( \{x_1\} \) are both treated as contributing the same value to the total sum. Normally, the four subsets involving these two elements---\( \emptyset \), \( \{x_0\} \), \( \{x_1\} \), and \( \{x_0, x_1\} \)---can produce up to four different partial sums. Under our rule, the sums for \( \{x_0\} \) and \( \{x_1\} \) become identical, collapsing these states into fewer unique outcomes.

To preserve correctness, the enumerator stores only a single \emph{canonical} representation for each resulting sum---typically the lexicographically minimal subset, encoded as a bitmask or ordered index list. This ensures that even when multiple subsets alias to the same sum due to injection, the representation is deterministic and non-redundant. This logic is consistent with the uniqueness guarantees discussed in Section~\ref{sec:enum}.

\subsection{Resulting Speed-up}

The \emph{Controlled Aliasing} technique allows us to reduce the number of explored configurations beyond the classical $O^*(2^{n/2})$ bound. By intentionally aliasing a constant number of elements on each side of the meet-in-the-middle split, and carefully pruning redundant states via canonical tracking, we reduce the size of the enumeration tree and accelerate pruning.

\medskip

\noindent \textbf{Theoretical Note.} A formal analysis of this strategy appears in the companion paper~\cite{ic-subsetsum}, which proves that BWSS, when equipped with controlled aliasing, achieves a worst-case runtime bounded by $O^*(2^{n/2 - \varepsilon})$ for some constant $\varepsilon > 0$. That argument relies on a combinatorial analysis of how injected duplicates collapse enumeration trees under controlled aliasing. We do not reproduce that proof here.

\medskip

\paragraph*{Empirical Effectiveness.}
Injecting one duplicate per half is often enough to cut the number of unique subset sums significantly in practice. If memory is tight, injecting two duplicates can yield further improvements at the cost of additional merge tracking. Each added duplicate trades memory and bookkeeping for reduced enumeration cost. A one-or-two duplicate configuration typically offers the best tradeoff.

\paragraph*{Half-Subset Enumeration with Aliasing.}
By default, we enumerate only subsets of size at most \(n/4\) in each half and rely on complements to recover the rest. This halves the enumeration cost to \(2^{n/2 - 1}\) per side. Injecting one collision per half reduces it further. In practice, combining these two optimizations reduces total enumeration cost to below 40\% of the naive baseline on average.

\paragraph*{Universality and Guaranteed Pruning.}
This Controlled Aliasing method works uniformly across all instances, whether structured or not. Once collisions are injected, the enumeration tree collapses early due to forced redundancy, and pruning becomes universal. Canonical tracking guarantees that the effective sumset remains correct and unique, even with injected duplicates.

\section{Divide and Conquer for Anytime Running Time}
\label{sec:divide_and_conquer}

In this section, we present a divide-and-conquer variant of our baseline solver \emph{without}
the Controlled Aliasing optimization. This choice lets us illustrate the incremental
and "resume" behavior more transparently. Our goal is to show how the algorithm can progressively
explore the problem space---whose effective size is determined by \( U = |\Sigma(S)| \) (the number of
unique subset sums)---in an anytime fashion. This variant also highlights how we leverage the
\textbf{seeding} and \textbf{branch enumeration} features of our enumerator.

\subsection{Overview of the Optimization}

Recall that our basic enumeration process expands \textbf{\(k\)-permutations} column by column to generate all feasible \emph{unique subset sums} for each split, thereby ultimately producing a set of distinct sums of size \( U \), which is typically much smaller than the full \(2^n\) possibilities. In addition to this standard expansion, our enumerator supports \emph{seeding}: if the initial \textbf{INPUT} contains partially developed \(k\)-permutations, their accumulated sums (which contribute to the overall \( U \)) are added to the memoization structure (sum \(\rightarrow\) canonical subset), and the enumeration resumes from these prefixes. This seeding capability ensures that intermediate results are retained and re-utilized, thereby connecting successive enumeration cycles while avoiding redundant work over the same unique sums.

\begin{figure}[H]
\centering
\includegraphics[width=1.0\textwidth]{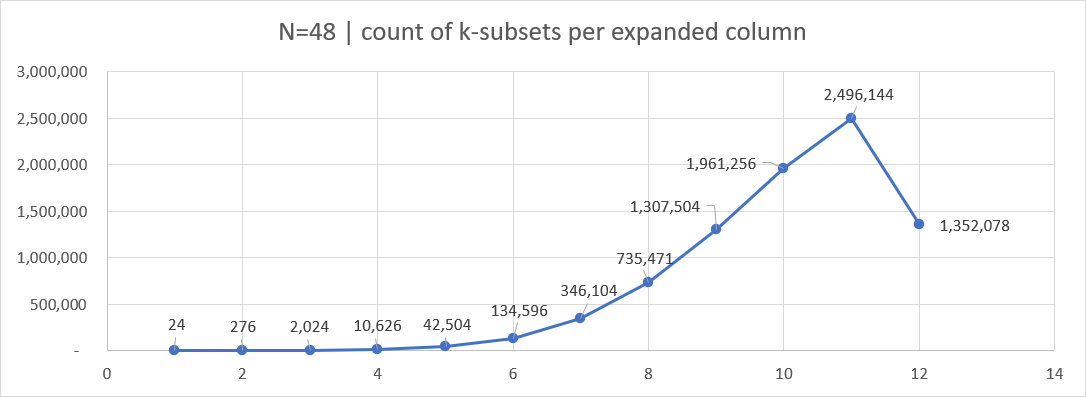}
\caption{Counts of \textbf{\(k\)-subsets} enumerated per column expansion (input \(n=48\)).}
\label{fig:regular}
\end{figure}

\textbf{Figure~\ref{fig:regular}} shows the number of \textbf{\(k\)-subsets} generated in each column expansion sub-process for an input instance with \(n=48\). In this case, the process is executed \(n/4=12\) times, yielding a total of \(8\,388\,607\) \textbf{\(k\)-subsets} per split, which corresponds to the expected \(\frac{2^{n/2}}{2}\) in the worst case. The memoization structure holds only half of the unique subset sums, reflecting the true effective search space.

\begin{figure}[H]
\centering
\includegraphics[width=1.0\textwidth]{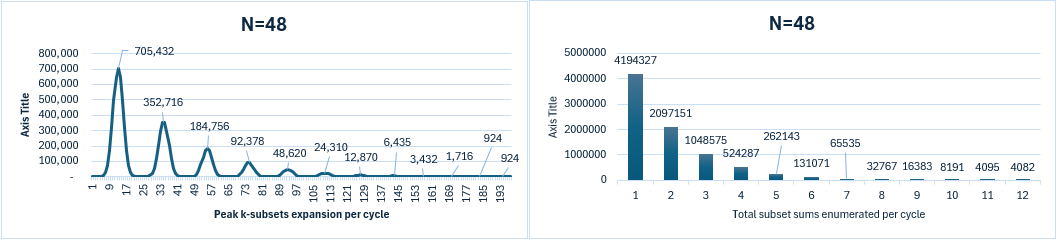}
\caption{Run-time graphs for \(n=48\) with optimization enabled. The left panel shows the most computationally expensive column expansion in each cycle, while the right panel illustrates the number of subsets enumerated in each cycle.}
\label{fig:optimized}
\end{figure}

Our improved approach modifies this behavior by slicing the enumeration into multiple cycles rather than performing one long cycle. Each cycle handles a different slice of the problem---updating and extending the memoized unique subset sums (up to size \( U \)) incrementally---as illustrated in Figure~\ref{fig:optimized} for the same \(n=48\) instance.

\subsection{The Look-Ahead and Rescheduling Mechanism}

Our optimization uses a look-ahead strategy to decide whether to postpone (reschedule) the expansion of a \textbf{\(k\)-permutation} to a later cycle. This strategy is customizable; the mechanism used in our experiments works as follows:

\begin{enumerate}
    \item For each \textbf{\(k\)-permutation} about to expand a new column, we examine a block of candidate elements. The size of this block is determined by:
    \[
    \mathsf{look\text{-}ahead} =
    \begin{cases}
    0, & \text{if } 2^n > \mathrm{SUM}(S), \\
    \left\lfloor \frac{n}{16} \right\rfloor +
    \begin{cases}
    \displaystyle \frac{n}{32} + 1, & \text{if } \frac{n}{32} > 1, \\
    0, & \text{otherwise,}
    \end{cases} & \text{otherwise.}
    \end{cases}
    \]

    \item During the \(i\)-th column expansion of the current cycle, we consider a block of candidate elements spanning from the \(i\)-th to the \((i+\mathsf{look\text{-}ahead})\)-th position. For each candidate not yet in the \(k\)-permutation, we compute the tentative new sum.

    \item If this tentative sum has already been registered in the memoization table (sum \(\rightarrow\) canonical subset), we defer the expansion to a future cycle:
    \[
    \mathsf{futureCycle} = \mathsf{currentCycle} + (\text{position of candidate in block})
    \]
\end{enumerate}

This mechanism prioritizes promising branches that may produce new canonical sums early. Rescheduled branches tend to yield fewer new entries. Memoized entries ensure continuity.

Each cycle processes about half the remaining work, forming a geometric decay. Denoting the first cycle’s work as \(W_0\), total work is:
\[
W_{\text{total}} = \sum_{i=0}^{\infty} W_i \approx 2W_0, \quad W_i \approx \frac{W_0}{2^i}
\]

\subsection{A Brief Example for Anytime/Online Usage}

Suppose the solver runs for two cycles on a 16-element split. You may pause it, inspect partial canonical subset sums, or insert a new element \(x_{\text{new}}\). The memo retains canonical state, so new branches resume without recomputation—exemplifying the \emph{anytime} nature.

\subsection{Experiments and Time Analysis for the Anytime Runtime}

\begin{figure}[H]
\centering
\includegraphics[width=1.0\textwidth]{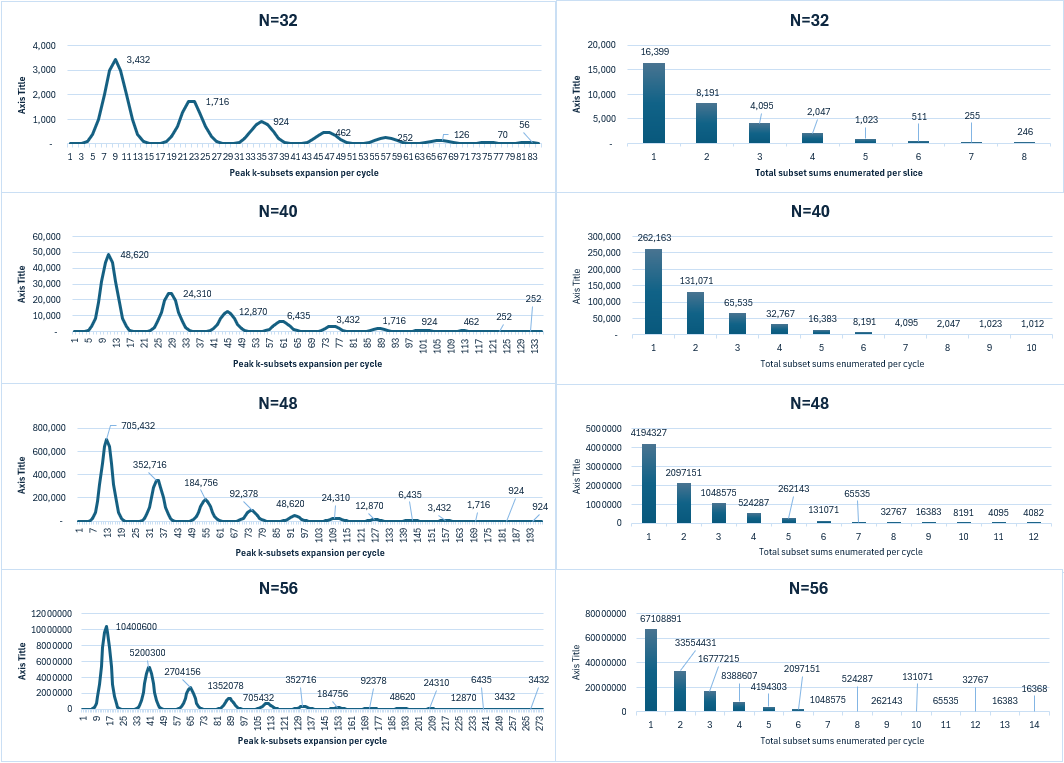}
\caption{Anytime run-time graphs for input instances of lengths 32, 40, 48, and 56.}
\label{fig:run_times}
\end{figure}

We conducted experiments on worst-case instances using a 12th Gen i7-12700 with 64GB RAM. Figure~\ref{fig:run_times} shows the subset counts and cycle times for \(n=32, 40, 48, 56\). First cycles dominate due to the prioritization heuristic.

\subsection{Intuition Behind the Slicing Mechanism}

The combinatorial tree exhibits fractal-like behavior: the topological pattern of partial sums persists across scales, making this layered expansion efficient and predictable.

\subsection{Initial Solution Time Complexity and Improvement Over Time Estimation}

The worst-case initial expansion (without controlle aliasing) is:
\[
O\left(\text{poly}(n) \cdot \binom{n/2 - \lfloor{n/16}\rfloor}{n/4 - \lfloor{n/32}\rfloor}\right)
\approx \widetilde{O}\!\left(\binom{n/2 - \lfloor{n/16}\rfloor}{n/4 - \lfloor{n/32}\rfloor}\right)
\]

Work per cycle decays geometrically:
\[
W_i \approx \frac{W_0}{2^i}, \quad i \le n/4
\]

\paragraph*{Note on Complexity.} While a full, formal analysis of the anytime variant's performance is beyond the scope of this paper, its efficiency is governed by the same principles as the core solver. The foundational complexity analysis of that core solver---establishing its adaptive runtime parameterized by $U$ and its strict worst-case bounds---is presented in a companion paper.

\section{Adaptive Time Complexity and Instance Hardness Analysis}
\label{sec:adaptivity}

A key advantage of our algorithm is its inherent adaptivity: its effective time complexity depends not only on the number of elements \( n \) and their bit-length \( w \) but also on the structure of the input as captured by \( U = |\Sigma(S)| \). In our approach, the interplay between dynamic programming and combinatorial enumeration naturally adjusts the computational effort based on the intrinsic characteristics of the input.

\subsection{Key Factors Influencing Instance Classification and Hardness}

Let \(n\) denote the number of elements in the set \(S\), and let \(w = \log_2(\max S)\) be the bit-length of the largest element. In addition to the classical measures of additive structure—such as the doubling constant \(C\) and additive energy \(E(S)\)—we also consider the extended structural parameters introduced earlier: the linearity factor \(\lambda\), the clustering factor \(\gamma\), and the duplicate measure \(\delta\). These parameters collectively capture the degree of additive redundancy in \(S\) and offer a finer-grained understanding of how collisions among subset sums occur.

We can broadly classify instances as follows:

\begin{itemize}
    \item \textbf{Dense Instances:} When \(w\) is relatively small compared to \(n\) (i.e., the elements are confined to a narrow numerical range) and the extended parameters indicate high redundancy (for example, low \(\lambda\) and \(\gamma\) or high \(\delta\)), many subsets yield identical sums. In these cases, the effective search space \(U = |\Sigma(S)|\) is dramatically smaller than \(2^n\).
    \item \textbf{Sparse Instances:} When \(w\) is large and the input exhibits little additive structure, the number of unique subset sums \(U\) approaches its worst-case bound.
    \item \textbf{Mixed Instances:} When \(S\) contains regions with differing structural properties, the overall behavior is a blend of the above cases.
\end{itemize}

For mixed instances, the input is partitioned at least into two subsets:
\begin{itemize}
    \item \( S_D \) (the dense part) with \( n_D \) elements and effective bit-length \( w_D \), leading to a smaller \( U_D \),
    \item \( S_S \) (the sparse part) with \( n_S \) elements and effective bit-length \( w_S \), where \( U_S \) is closer to \(2^{n/2}\).
\end{itemize}
Naturally, \( n_D + n_S = n \).

\subsection{Topological vs. Numerical Target Perspective}

In a classical sense, one might guess that \( \Sigma(S)/2 \) is the hardest target because enumerating all subsets up to that value can be combinatorially explosive. However, our column-by-column enumerator is structured so that \emph{each level} of expansion focuses on all subsets (permutations) of size \( k \) \emph{before} moving on to size \( k+1 \). Consequently, partial sums near \( \Sigma(S)/2 \) may be discovered (and pruned, if duplicated) relatively early—often via complementary checks around half of \( \ell_0 \) and half of \( \ell_1 \).

By contrast, a target \( t \) that \emph{requires} (for instance) \emph{the maximum \( k \)-subset expansions} (i.e., near \( k = n/4 \) per split) is ``topologically deeper'' in the enumerator’s tree. The solver only finalizes those largest subsets after it has exhausted all smaller-\( k \) expansions. Hence, from the enumerator’s standpoint, \emph{the hardest target is not numerically about being near \( \Sigma(S)/2 \), but rather about being reached only by large or late-expanding subsets}.

\subsection{Output-Sensitive Perspective}

While our solver does not enumerate the full sumset \(\Sigma(S)\), its core components---the partial enumerators for \(\ell_0\) and \(\ell_1\)---are output-sensitive: each runs in time proportional to the number of unique subset sums it produces. If the goal is to explicitly list all such sums from either half, then the runtime is necessarily \emph{optimal} from an output-sensitive standpoint.

More generally, any algorithm that fully constructs \(\Sigma(S)\) must incur at least \(O(U)\) time, where \(U = |\Sigma(S)|\). However, our double meet-in-the-middle approach avoids this cost by only generating the partial sumsets and checking for compatible pairs on-the-fly. As such, it is not output-sensitive in \(U\), but it is substantially faster than \(O(U)\) in many cases. The tradeoff is deliberate: we achieve sublinear-in-\(U\) decision time by not constructing the full output.

\paragraph*{Remark on Stability.}  
Because each sum is retained with a fixed canonical subset, the enumerator’s output is stable across runs and supports reproducible debugging or downstream synthesis tasks.

\subsection{Comparison to Dynamic Programming}
\label{sec:dp_comparison}

Dynamic programming (DP), notably Bellman’s algorithm, solves \textsc{Subset Sum} in time $O^*(n \cdot t)$, where $t$ is the numeric target. This method is numerically sensitive but structurally agnostic: it assumes all values from $1$ to $t$ may be reachable and allocates effort accordingly. As such, DP is well-suited for instances with small numeric magnitude but cannot exploit redundancy or structure in the input.

In contrast, our algorithm achieves a certificate-sensitive runtime of $O^*(U \cdot n^2)$ deterministically (or $O^*(U \cdot n)$ randomized), where $U = |\Sigma(S)|$ denotes the number of unique  subset sums induced by the input. This parameter captures the effective search space: the number of structurally distinct states the algorithm must consider. 

\begin{itemize}
    \item When $U \ll t$, as is often the case in compressible or structured instances, our algorithm achieves asymptotic speedups over DP by avoiding unnecessary work on unreachable sums.
    \item When $U \approx t$, the algorithm remains competitive, matching the performance of DP up to a polynomial overhead in $n$.
\end{itemize}

No improvement to DP that depends on $t$ can dominate our algorithm across all inputs. Since $U$ is instance-dependent and often exponentially smaller than $t$, the algorithm redefines tractability for large-scale \textsc{Subset Sum} instances. Its runtime is governed not by numeric size but by structural entropy—making it adaptively faster where classical methods are inherently blind.

\section{Adaptive Runtime Experiments}\label{sec:runtime_experiments}

We perform experiments to demonstrate how the cost of generating \( U \) depends on various structural properties, and how the adaptive enumeration model self-adjusts during computation to optimally construct the specific combinatorial tree of the input instance.

All experiments were executed using the non-anytime version of the algorithm to simplify the runtime analysis to a single cycle. Additionally, solution finding was disabled to simulate the worst-case scenario—a no-solution instance that requires exhaustive exploration of the problem space.

\subsection{Dense Instances}

In this experiment, we generate a dissociative (collision-free) instance with \(n=48\) and \(w=48\), yielding \(U = 2^{n/2}\) effective unique subset sums. In this baseline instance, all elements are distinct and sufficiently spread out so that no additional additive structure is exploited.

To ensure the accuracy of the experiment, every element is confirmed to be exactly \(w\) bits long. This precaution avoids creating a mixed instance, where certain regions of the input might be denser than others.

\begin{figure}[H]
\centering
\includegraphics[width=1.0\textwidth]{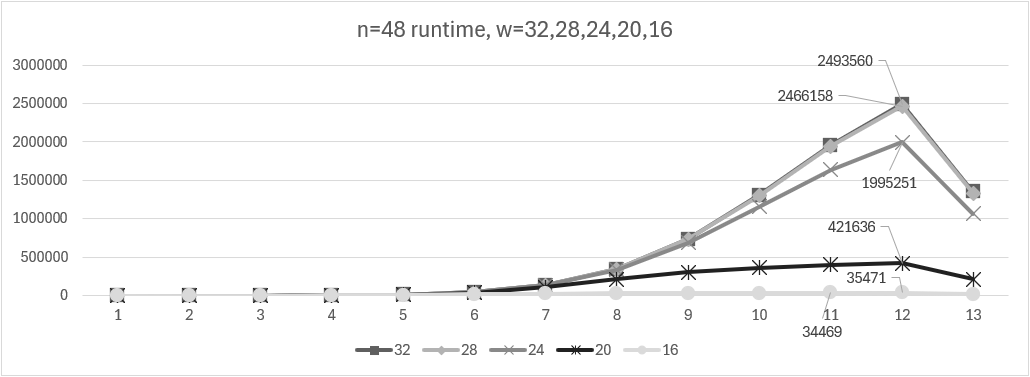}
\caption{Counts of \textbf{\(k\)-subsets} enumerated per column expansion for a (input \(n=48\)) input instance with elements of varying bit-length from w=32 to w=20 decreasing in steps of 4 bits.}
\label{fig:n48runtime_with_densities}
\end{figure}

Figure~\ref{fig:n48runtime_with_densities} illustrates the effect of density on the enumeration process. We identify a threshold at approximately \(w \approx 32\) (corresponding to a density of 1.5), below which density begins to influence the total number of unique subset sums progressively. We then reduce the bit-length of the elements by 4 bits at each step (effectively shifting the numbers right) to generate instances with higher density. This setup demonstrates the progressive runtime dynamism of the enumerator as the effective search space \(U\) is reduced.\\

\begin{table}[H]
\centering
\begin{tabular}{l c r r}
\hline
\textbf{Scenario} & \textbf{Density} & \textbf{Unique Subset Sums} & \textbf{Diff with Previous} \\
\hline
baseline (n=48/w=48)& 1.0 &  8\,388\,607 & -- \\
n=48/w=32 & 1.50 & 8,382,135 & 99.92\% \\
n=48/w=28 & 1.71 & 8,311,785 & 99.16\% \\
n=48/w=24 & 2.00 & 7,061,331 & 84.96\% \\
n=48/w=20 & 2.40 & 2,072,581 & 29.35\% \\
n=48/w=16 & 3.00 & 227,034  & 10.95\% \\
\hline
\end{tabular}
\caption{Comparison of Unique Subset Sums and Density for $n=48$ with varying  bit-length $w$. Each unique sum is represented canonically.}
\label{tab:subset-sums-density}
\end{table}

\paragraph*{Transition Phase Analysis and Time Complexity Approximation.} 
Let 
\[
w = \log_2\Bigl(\max_{x\in S} x\Bigr)
\]
be the bit-length of the elements in the input set \(S\). In the worst-case scenario—when no additive structure is exploited—the classical meet-in-the-middle algorithm generates roughly 
\[
2^{n/2}
\]
unique subset sums. However, if the elements in \(S\) are small (i.e., \(w\) is small), then the total sum of all elements is upper bounded by
\[
\sum_{x\in S} x = O\Bigl(n \cdot 2^w\Bigr)
\]
By the pigeonhole principle, even if we consider all possible subsets, the number of unique subset sums (i.e., the effective search space) is at most
\[
U = O\Bigl(n \cdot 2^w\Bigr)
\]

Thus, when the input is dense (i.e., when the range of possible sums is much smaller than the number of candidate subsets), many different \(k\)-subsets yield the same sum. This phenomenon occurs when
\[
n \cdot 2^w \ll 2^{n/2}
\]
Taking logarithms of both sides gives
\[
w < \frac{n}{2} - \log_2 n
\]
In other words, if the bit-length \(w\) of the input elements satisfies
\[
w < \frac{n}{2} - \log_2 n
\]
then the effective search space \(U\) is bounded by
\[
U = O\Bigl(n \cdot 2^w\Bigr)
\]
which is exponentially smaller than \(2^{n/2}\). 

This threshold marks the \emph{transition phase} of the enumerator: when \(w\) is below this threshold, the algorithm benefits from a high degree of collisions (i.e., many \(k\)-subsets yield the same sum), resulting in a dramatic reduction in the number of branches that must be explored as the density increases. Furthermore, since our enumeration model implements a double meet-in-the-middle strategy, the maximum number of iterations is \(n/4\) (as observed in our experimental runtime graphs). 

Consequently, by substituting this bound for $U$ into the general runtime $O^*(U \cdot n^2)$, we obtain:

\[
O(U \cdot n^2) = O(n^2 \cdot n \cdot 2^w) = O(n^3 \cdot 2^w),
\]

which is pseudopolynomial in $w$ and strictly better than the worst-case $O^*(2^{n/2})$ bound when \(w < \frac{n}{2} - \log_2 n\).

This reflects the deterministic case; a randomized variant using hashing would reduce the complexity to $O^*(n^2 \cdot 2^w)$ in this regime.

\subsection{Extremely Dense Instances}

In this experiment, we process several instances with \( n = 100 \) and varying \( w \) values (16, 20, and 24), with \( d > 3.5 \).

In extremely dense instances, the maximum average expansion per element is \( 2^w \), and the maximum number of column expansions is \( n/4 \).

\begin{figure}[H]
\centering
\includegraphics[width=1.0\textwidth]{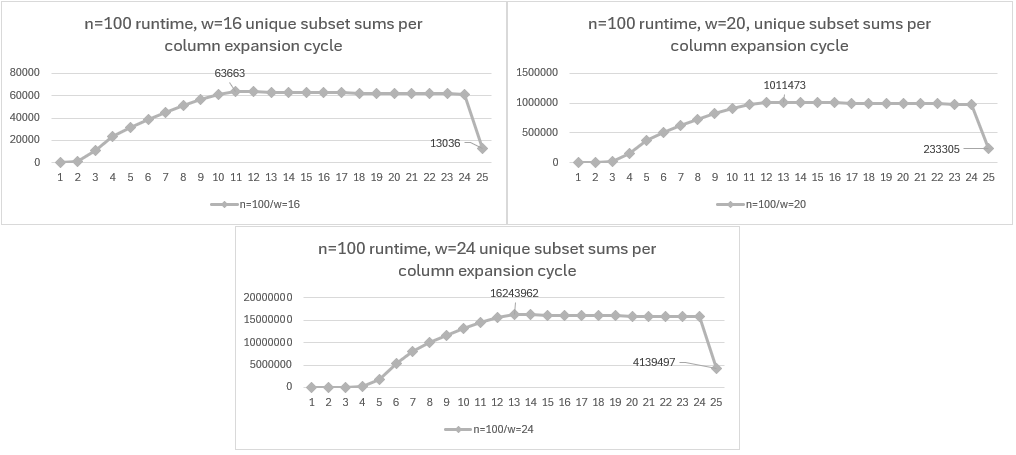}
\caption{Counts of \textbf{\(k\)-subsets} enumerated per column expansion (input \( n = 100 \) with \( w = 16, 20, 24 \)).}
\label{fig:n48experiments_highdensity}
\end{figure}

In extremely dense instances, the pigeonhole principle forces so many collisions that the enumeration must expand through all required columns, resulting in a runtime dominated by these dense expansions. This outcome is precisely what one would expect from the inherent column-by-column dynamics of our approach.

\subsection{Duplicates Elements}

\begin{figure}[H]
\centering
\includegraphics[width=1.0\textwidth]{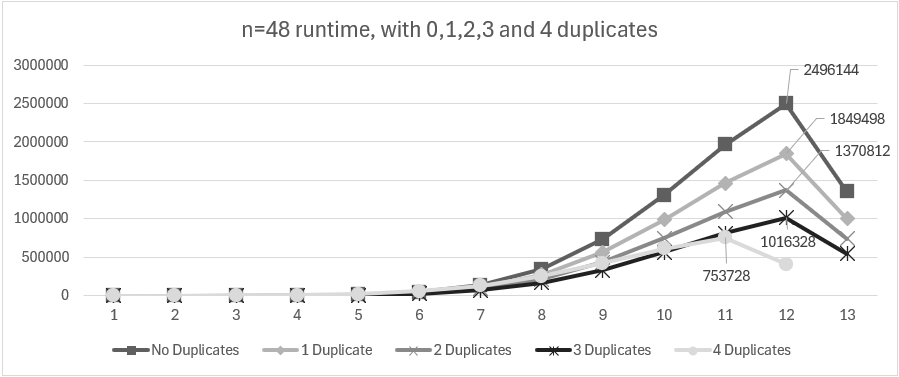}
\caption{Counts of \textbf{\(k\)-subsets} enumerated per column expansion (input \(n=48\)).}
\label{fig:n48runtime_with_duplicates}
\end{figure}

Figure~\ref{fig:n48runtime_with_duplicates} illustrates the effect of duplicates during enumeration. We start with a dissociative instance (i.e., no duplicates) and progressively introduce duplicates (up to 4 per split), duplicating always a different element.

We observe that the most expensive enumeration matches our baseline results, and that each additional duplicate reduces the number of effective unique subset sums by \(~25\%\) from the previous run. These sums are tracked using canonical bitmask representations, avoiding redundant credit for distinct but permutationally equivalent subsets.

\begin{table}[H]
\centering
\begin{tabular}{l|r|c}
\hline
\textbf{Scenario} & \textbf{Unique Subset Sums} & \textbf{Diff. with Previous} \\
\hline
No Duplicates  & 8\,388\,607 & -- \\
1 Duplicate    & 6\,291\,455 & 75\% \\
2 Duplicates   & 4\,718\,591 & 75\% \\
3 Duplicates   & 3\,538\,943 & 75\% \\
4 Duplicates   & 2\,654\,207 & 75\% \\
\hline
\end{tabular}
\caption{Number of effective unique subset sums (canonical) for various levels of duplication.}
\label{tab:duplicates}
\end{table}

\paragraph*{Duplicates Approximate Time Complexity Analysis}

The experimental data strongly suggests an approximate complexity scaling for instances with \( \delta \) duplicates. As shown in Table~\ref{tab:duplicates}, each additional duplicate reduces the number of unique subset sums by a consistent factor of approximately 0.75. This implies that for an unstructured instance that would otherwise produce roughly \( 2^{n/2} \) distinct sums, the introduction of \( \delta \) duplicates reduces this count to approximately
\[
U(\delta) \approx 2^{n/2} \cdot (0.75)^\delta.
\]
This exponential reduction in the size of the effective search space \( U \) directly translates to significant performance gains, illustrating the solver’s ability to exploit structural redundancy. A full theoretical derivation of the resulting worst-case exponent reduction is deferred to a companion paper.

\subsection{Additive Structure}

In this experiment, we process an instance with \(n=48\) and introduce arithmetic progressions to observe the effect of additive structure on the number of unique subset sums. First, we insert one arithmetic progression (a sequence) of 3 elements and then extend it to 4 elements. Next, we introduce a disjoint progression to form two sequences—first with 3 elements each, and then extend both sequences to 4 elements each. This experimental design allows us to verify the effect of having one versus two sequences per split, as well as the impact of increasing the sequence length, on the reduction of unique subset sums.

\begin{figure}[ht]
\centering
\includegraphics[width=1.0\textwidth]{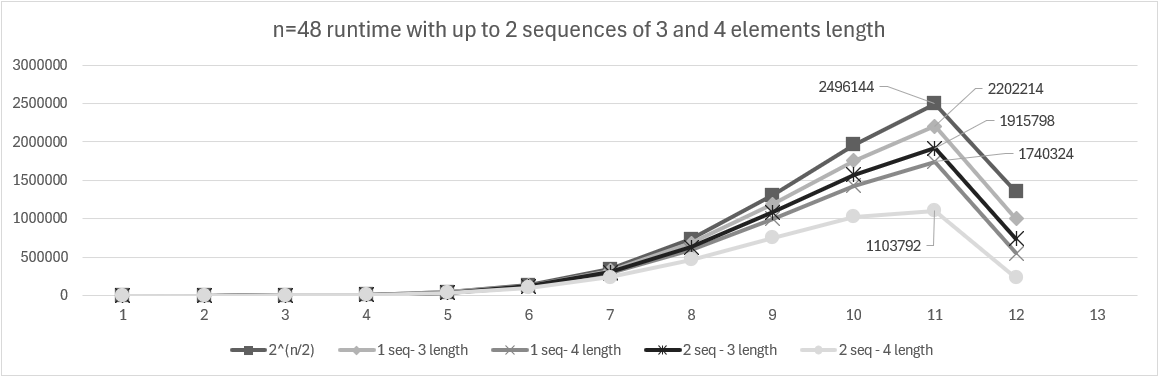}
\caption{Counts of \textbf{\(k\)-subsets} enumerated per column expansion (input \(n=48\)) when the input instance includes 1 or 2 arithmetic progressions of 3 and 4 elements.}
\label{fig:n48experiment_additivestructure}
\end{figure}

\begin{table}[H]
\centering
\begin{tabular}{l c c}
\hline
\textbf{Scenario} & \textbf{Unique Subset Sums} & \textbf{Diff with Previous} \\
\hline
No sequences (\(2^{n/2}\)) & 8\,388\,607 & -- \\
1 seq -- 3 length & 7\,340\,031 & 87.50\% \\
1 seq -- 4 length & 5\,767\,167 & 78.57\% \\
\hline
\hline
No sequences (\(2^{n/2}\)) & 8\,388\,607 & -- \\
2 seq -- 3 length & 6\,422\,527 & 76.56\% \\
2 seq -- 4 length & 3\,964\,927 & 61.73\% \\
\hline

\end{tabular}
\caption{Unique Subset Sums and percentage difference for the one and two sequence cases.}
\label{tab:one_sequence}
\end{table}

We observe that the presence of two sequences reduces the number of unique subset sums more than a single sequence does. Furthermore, extending the sequence length by one element results in an additional reduction of approximately 9\% for the one-sequence case and about 15\% for the two-sequence case. This demonstrates that the effect of additive structure compounds, thereby reducing the total number of unique subset sums.

Additional experiments with other types of sequences (e.g., arithmetic progressions, quadratic sequences, and certain Fibonacci-like sequences) confirm that structured sequences with near-linear or progression-like behavior can significantly reduce the total number of unique subset sums below \(2^{n/2}\)..

\subsection{Summary}
The conducted experiments demonstrates that adaptive behavior of our algorithm is directly tied to the effective search space \( U \) which is dependent on the structural characteristics of the input instance. In particular, the overall runtime improves when the input instance exhibits structural characteristics that reduce \( U \) relative to the worst-case bound. For example, when the instance is:

\begin{itemize}
  \item \textbf{Dense:} When the elements are confined to a narrow numerical range (i.e., small bit-length \( w \)), many subset sums collide, yielding a significantly reduced \( U \). In this regime, the deterministic variant of our solver achieves a pseudopolynomial runtime of \(O(n^3 \cdot 2^w)\), reflecting the cost of canonical bitmask comparisons. A randomized variant reduces this to \(O(n^2 \cdot 2^w)\) by using hash-based memoization, matching the classical dynamic programming bound \(O(n \cdot t)\) when \( t = O(n \cdot 2^w) \), but with substantially less redundant computation.

    \item \textbf{Redundant (Duplicates):} The presence of duplicate elements inherently reduces the number of unique subset sums by eliminating redundant branches.
    \item \textbf{Structured (Linearity and Clustering):} When the elements exhibit near-linear relationships or are tightly clustered, the resulting additive structure forces many \(k\)-subsets to produce the same sum, thereby reducing \( U \).
\end{itemize}

In contrast, in sparse instances where the elements are widely spread and lack significant structure, \( U \) remains close to the worst-case \(2^{n/2}\), and the algorithm reverts to near-worst-case performance. For mixed instances, the overall runtime reflects a combination of these behaviors. This refined understanding of instance hardness—captured by \( U \) and influenced by factors such as duplicates, linearity, and clustering—enables our solver to gracefully adapt to the specific characteristics of the input without requiring explicit control flow adjustments, specialized pre-processing, or prior knowledge of the instance's structure.

\subsection*{Concluding Remark on Adaptive Hardness.}
Our experimental and theoretical analysis confirms that the effective search space \(U = |\Sigma(S)|\) is intricately governed by the input's additive structure—whether measured through small doubling, high additive energy, specialized sequence patterns, or duplicate redundancy. These combinatorial properties directly dictate the number of unique subset sums that our algorithm must process, and thus, they are the key to its adaptive pruning strategy and overall speedup. This unified perspective not only explains the observed transition from worst-case to near-dynamic programming performance in structured instances but also provides a clear roadmap for further optimizing NP-complete solvers by focusing on the inherent structure of the input.

\section{Comparison with Existing Solvers}
\label{sec:comparision_with_solvers}

Table~\ref{tab:comparison_solvers} summarizes how traditional state-of-the-art approaches compare with the proposed adaptive, structure-aware solver.

\begin{table}[H]
\centering
\small
\setlength{\tabcolsep}{4.5pt}
\renewcommand{\arraystretch}{1.25}
\begin{tabular}{p{2.6cm}|p{2.2cm}|p{1.5cm}|p{1.7cm}|p{1.6cm}|p{2.2cm}|p{2.0cm}}
\hline
\textbf{\scriptsize Algorithm / Approach} 
& \textbf{\scriptsize Time Complexity} 
& \textbf{\scriptsize Memory} 
& \textbf{\scriptsize Uses Structure?} 
& \textbf{\scriptsize Anytime?} 
& \textbf{\scriptsize Key Characteristics} 
& \textbf{\scriptsize Limitations} 
\\ \hline

\scriptsize Dynamic Prog. (Bellman)~\cite{bellman-1954} 
& \scriptsize $O(n \cdot t)$ 
& \scriptsize $O(t)$ 
& \scriptsize No 
& \scriptsize No 
& \scriptsize Exact, simple, pseudo-poly 
& \scriptsize Inefficient when $t$ large 
\\ \hline

\scriptsize Koiliaris--Xu (2017)~\cite{koiliaris-xu-2017} 
& \scriptsize $\widetilde{O}(t \sqrt{n})$ 
& \scriptsize $O(t)$ 
& \scriptsize Yes (algebraic; not adaptive) 
& \scriptsize No 
& \scriptsize Deterministic; FFT-based for $t$-bounded regime 
& \scriptsize Pseudo-poly; not output-sensitive 
\\ \hline

\scriptsize Classical MITM (Horowitz--Sahni)~\cite{horowitz-sahni-1974} 
& \scriptsize $O^*(2^{n/2})$ 
& \scriptsize $O^*(2^{n/2})$ 
& \scriptsize No 
& \scriptsize No 
& \scriptsize Deterministic split + merge 
& \scriptsize High time and space; blind to structure 
\\ \hline

\scriptsize Schroeppel--Shamir (1979)~\cite{schroeppel-shamir-1981} 
& \scriptsize $O^*(2^{n/2})$ 
& \scriptsize $O^*(2^{n/4})$ 
& \scriptsize No 
& \scriptsize No 
& \scriptsize Space-optimized MITM variant 
& \scriptsize Still exponential in time 
\\ \hline

\scriptsize Howgrave--Graham--Joux (2010)~\cite{howgrave-graham-joux-2010} 
& \scriptsize avg. $O^*(2^{n/2 - \delta})$ 
& \scriptsize $O^*(2^{n/2})$ 
& \scriptsize Partial (random instance assumptions) 
& \scriptsize No 
& \scriptsize Dissection methods; good on average-case 
& \scriptsize No worst-case guarantee; structure-agnostic 
\\ \hline

\scriptsize Bringmann--Fischer--Nakos (2025)~\cite{bringmann2024subsetsumunconditional} 
& \scriptsize $\widetilde{O}(|\mathcal{S}(X,t)| \cdot \sqrt{n})$ 
& \scriptsize $\widetilde{O}(|\mathcal{S}(X,t)|)$ 
& \scriptsize Yes (additive combinatorics, sumset bounds) 
& \scriptsize Yes 
& \scriptsize Las Vegas; output-sensitive; algebraic convolution 
& \scriptsize Randomized; derandomization open 
\\ \hline

\scriptsize Meet-in-the-middle refinements~\cite{nederlof2021improving} 
& \scriptsize $O^*(2^{n/2})$ 
& \scriptsize $O^*(2^{n/4})$ 
& \scriptsize Partial (space/time optimizations) 
& \scriptsize No 
& \scriptsize Deterministic refinements of classical MITM 
& \scriptsize Still exponential; not structure-aware 
\\ \hline

\scriptsize \textbf{This Work} 
& \scriptsize $O(U \cdot n^2)$ deterministic, $\mathbb{E}[O(U \cdot n)]$ randomized 
& \scriptsize $O(U \cdot n)$ 
& \scriptsize \textbf{Yes (full structure-aware)} 
& \scriptsize \textbf{Yes} 
& \scriptsize Adaptive Enumeration, U-sensitive Pruning, Canonical Pruning, Incremental
& \scriptsize Exponential if $U \approx 2^{n/2}$ 
\\ \hline

\end{tabular}
\caption{Comparison of existing Subset Sum solvers with our adaptive, structure-aware algorithm.}
\label{tab:comparison_solvers}
\end{table}

\section{Key Insights and Benefits}\label{sec:additional_benefits}

Beyond the core complexity guarantees and adaptivity to \( U = |\Sigma(S)| \) discussed above, our algorithm exhibits several further properties that enhance its practical and theoretical appeal. Many of these benefits stem from our focus on the effective search space—namely, the unique subset sums, whose total number is \( U = |\Sigma(S)| \). Throughout, we store only \emph{canonical representations} of these sums, ensuring that no redundant paths are explored. We summarize these properties below.

\subsection*{Anytime and Incremental Behavior}
Our algorithm’s slicing and rescheduling techniques enable it to produce intermediate results during the enumeration process. This anytime behavior is particularly valuable in time-critical or resource-constrained environments, as it allows the algorithm to be interrupted at any point while still providing partial, useful outputs and the exact solution upon full execution. The incremental nature of the combinatorial tree exploration, which tracks only canonical representations of unique subset sums (i.e., up to \( U \) distinct sums), ensures that progress is steadily made toward a solution.

\subsection{Online Updates as a New Branch}\label{sec:online_as_branch}
Beyond the anytime behavior described in Section~\ref{sec:divide_and_conquer}, our solver can also operate \emph{online}, handling newly arrived elements without discarding previous partial enumerations. In practice, adding a new element \(x_{\text{new}}\) to one of the splits (\(\ell_0\) or \(\ell_1\)) is treated exactly like introducing a new “branch” in the enumerator:
\begin{enumerate}
  \item We record \(x_{\text{new}}\) in the relevant split's data structure and update the total sum of that split accordingly.
  \item The algorithm’s standard column-expansion process (including any look-ahead or slicing rules) then applies to \(x_{\text{new}}\) just as it would to any other element. Consequently, \(x_{\text{new}}\) will contribute to new \(k\)-permutations, potentially introducing additional unique subset sums. Importantly, the memoization mechanism continues to track all previously generated unique sums as \emph{canonical representations}, avoiding duplicates and preserving correctness even under dynamic changes.
\end{enumerate}
This property makes the solver suitable for real-time or streaming variants of Subset Sum, where elements are revealed incrementally or dynamically added during computation.

\subsection*{Adaptive Prioritization and Governability}
A distinctive benefit of our anytime exploration is that it renders the search process highly governable. Since the algorithm operates in configurable discrete cycles—each producing a set of candidate branches that contribute to \( U \)—we can reassess the search state after every cycle. Once a cycle completes, the scheduler can reorganize the rescheduled branches by prioritizing those likely to yield new unique \emph{canonical} subset sums, pruning those that are redundant (i.e., those whose sums already exist in \( \Sigma(S) \)), and even deliberately encouraging collisions to block redundant exploration in unpromising regions of the search space. This adaptive prioritization not only guarantees the delivery of partial solutions at any interruption point but also enables the solver to be controlled and fine-tuned in real time, enhancing overall efficiency.

\subsection*{Early Capture of Imbalanced Solutions}
Any solution that requires an imbalanced distribution of elements between the two splits is typically detected earlier in the enumeration process. In our algorithm, the worst-case scenario arises when $\approx$ \(n/4\) elements are required from at least one split. In contrast, if a valid solution exists that uses fewer than \(n/4\) elements from one split, it will be found in an earlier cycle. The more imbalanced the solution (i.e., the greater the difference in the number of elements chosen from each split), the sooner its corresponding unique canonical subset sum is generated, leading to faster detection of a solution. This is because small subsets are explored early in each split, and canonical ordering ensures no duplicate effort across cycles.

\subsection*{Potential for Integration with Other Heuristics}
The structural design of our algorithm facilitates the integration of additional heuristics or approximation techniques. In scenarios where an exact solution is not strictly necessary, heuristic methods may be incorporated into the combinatorial tree exploration or the dynamic programming phase to further accelerate computation. By leveraging the reduced effective search space \( U \), such hybrid approaches can balance exactness with speed, broadening the practical utility of our framework.

\subsection*{Approximation via Bit Clearing and Rounding Error Bound}
One further advantage of our approach is the ability to trade a small amount of accuracy for significant efficiency gains. By clearing a fixed number of bits from the input values and the target, we reduce the effective bit-length, effectively pushing the problem into a denser regime where collisions (and hence a smaller \( U \)) occur more frequently. Although this introduces a bounded rounding error, it substantially reduces the computational load without compromising overall accuracy, and this effect is amplified under canonical pruning.

\subsection*{Parallelizability}
The inherent structure of our algorithm naturally lends itself to parallel processing. The division of the input into two independent splits (via the double meet-in-the-middle approach) and the independent processing of different slices of the combinatorial tree create multiple subproblems that can be handled concurrently. As each subproblem is focused on generating a portion of the unique subset sums (contributing to \( U \)), the workload can be distributed efficiently across multiple cores or nodes, further reducing practical running times. This parallel structure aligns well with modern multi-core architectures, and could further be accelerated on GPUs or distributed systems with minimal coordination overhead due to the independence of enumerated branches.

\paragraph*{Versatility for Subset Sum Variants.} The principles of this solver---enumerating unique states and using canonical representations for pruning---provide a versatile approach that could be adapted for closely related problems. For example, variants like the Multiple Subset Sum Problem or certain Knapsack problems, where solution states can also exhibit high collision rates, might benefit from a similar structure-aware enumeration strategy. This suggests a promising direction for applying these techniques beyond the classical problem formulation.

\subsection{Space Complexity}\label{sec:space-complexity}

Classical meet-in-the-middle algorithms (e.g., Horowitz--Sahni) require \( O(2^{n/2}) \) memory to store all partial subset sums of each half. In contrast, our algorithm stores only the \emph{canonical encodings} of the unique subset sums, whose effective count is \( U = |\Sigma(S)| \). Each encoding can be represented as a bitmask of length \( n \), yielding a total space usage of at most \( O(U \cdot n) \) bits.

This provides substantial memory savings on structured instances where \( U \ll 2^{n/2} \). 

\section{Emergent Dynamism and Instance Hardness}\label{sec:emergence}

A central feature of our algorithm is its ability to dynamically adapt to the structure of the input by focusing on the effective search space—namely, the number of \emph{unique subset sums} \(U = |\Sigma(S)|\). Without requiring any global knowledge of the full combinatorial tree, the enumerator incrementally extends partial \(k\)-permutations only when the resulting \(k\)-subset sum has not been encountered before. This pruning is guided by the canonical representation of subsets, which ensures that \emph{only one} path per sum is retained, preserving correctness while reducing enumeration.

\subsection{Instance Hardness Classification}

The effective complexity of a given instance is governed by the value of \(U\), rather than the total number of possible subsets. In practice:

\begin{itemize}
  \item \textbf{Hard Instances} are those for which \(U\) approaches the worst-case \(2^{n/2}\), necessitating near-exhaustive exploration. However, even in these cases, the use of \emph{injected collisions} (as described in Section~\ref{sec:duplicate-speedup})—via one or two syntactically aliased elements—can compress the combinatorial tree structure. This guarantees that enumeration remains strictly sub-\(2^{n/2}\), even when no structural redundancy is present in the input.

  \item \textbf{Structured Instances} (or sub-\(2^{n/2}\) instances) exhibit natural additive redundancy—through duplicate elements, near-linear arrangements, or clustering of values—that causes many subsets to map to the same sum. In these cases, the value of \(U\) is significantly smaller than \(2^{n/2}\), and the solver prunes large fractions of the tree without requiring injected collisions.
\end{itemize}

\noindent
Thus, while the worst-case bound \(O^*(2^{n/2})\) corresponds to fully unstructured inputs, most real-world or structured instances—and indeed any instance where collisions are syntactically injected—exhibit \(U \ll 2^{n/2}\), leading to considerably reduced enumeration and runtime.

\subsection{Scaling Invariance}

A notable property of our algorithm is that the enumeration cost, as measured by \(U\), remains invariant under uniform scaling of the input elements and the target. That is, multiplying all values by a fixed constant does not alter the number of unique subset sums. This emphasizes that instance hardness arises from the \emph{combinatorial structure} of the input, not from its absolute numerical values. Hence, \(U\) is governed by additive collisions and structural redundancy rather than magnitude.

\section{Future Directions}

The adaptive framework of the solver presented in this paper opens several avenues for future work focused on practical performance and algorithmic enhancement.

\begin{itemize}
  \item \textbf{Real-Time Instance Profiling:} The collision rate observed by the enumerator provides a real-time signal of an instance's structural redundancy. Future work could develop this into a robust heuristic for online instance classification, allowing a hybrid solver to choose the best strategy for a given input.
  
  \item \textbf{Enhanced Data Structures and Parallelization:} The performance of the solver hinges on the efficiency of its memoization table. Exploring more cache-efficient, succinct, or concurrent data structures could yield significant practical speedups. Furthermore, the algorithm's design is naturally suited for parallel deployment, and a full analysis of its scalability is a promising direction.
  
  \item \textbf{Heuristic-Augmented Search:} While our solver is deterministic, its governable, cycle-by-cycle enumeration process could be augmented with heuristics. For example, search could be prioritized toward branches that are heuristically more likely to contain the target, potentially finding a solution faster in applications where only one solution is needed.
  
  \item \textbf{Application to Subset Sum Variants:} A key open question is whether the principle of tracking unique state representations can be effectively applied to other closely related problems. Variants like the Multiple Subset Sum Problem or specific types of Knapsack problems, where solution states also exhibit high collision rates, are natural candidates for exploration.
\end{itemize}

\section{Conclusions}

We have introduced a novel solver for the Subset Sum problem that improves upon the classical meet-in-the-middle algorithm through a structure-aware, adaptive enumeration framework. Our approach pivots from traditional worst-case analysis to a focus on the effective search space $U = |\Sigma(S)|$---the number of unique subset sums. In structured instances where this space is exponentially smaller than the worst-case $2^n$, our solver achieves substantial speedups.

The key benefits of this algorithmic framework include:
\begin{itemize}
    \item \textbf{Instance-Aware Adaptivity:} The solver dynamically adjusts its work to the input's structure, with performance governed by $U$, not just $n$.
    \item \textbf{Reduced Enumeration:} By enumerating only subsets of size up to $n/4$ per half and using complements, we halve the number of required subset evaluations.
    \item \textbf{Sorting-Free Matching:} Each new unique sum is checked for a solution via a constant-time hash lookup, avoiding the expensive sorting and merging phases of classical MIM.
    \item \textbf{Anytime and Online Operation:} The solver's design allows for incremental computation, meaning it can be paused, resumed, and updated with new elements without restarting.
\end{itemize}

A central contribution of this work is the \textit{Controlled Aliasing} technique. By syntactically forcing collisions within the enumeration logic, our solver guarantees a strict, constant-factor reduction in the search space. This ensures that even for adversarial, unstructured inputs, the enumeration work is provably less than that of naive meet-in-the-middle, achieving a sub-$2^{n/2}$ worst-case bound.

In summary, this paper provides the complete blueprint for a deterministic solver that is not only adaptive to instance structure but is guaranteed to be faster than the classical MIM approach on all inputs. The broader theoretical implications of this framework, including its formal connection to Instance Complexity, are explored in our companion papers.

\bibliography{main}
\newpage
\appendix

\section{Unique Subset Sums Enumerator Algorithm}
\label{app:algorithms}

\begin{algorithm}
\label{algo:enumerator}
\caption{Unique Subset Sums Enumerator (Canonical-Aware)}
\begin{algorithmic}[1]
\begin{scriptsize}

\State \textbf{struct} \texttt{kPerm(sum, elements)} \{ \Comment Canonical k-permutation
\State \quad \texttt{sum} \(\gets 0\) \Comment Current subset sum
\State \quad \texttt{elements[n]} \(\gets \{0,0,\ldots\}\) \Comment Inclusion bitmask
\State \}
\State
\State \(S \gets \{\ldots\}\)
\State \(n \gets \text{len}(S)\)
\State \(MemoizedSums \gets \{\}\)
\State \(CanonicalSubsets \gets \{\}\)
\State \(Input \gets \{\texttt{kPerm}(0, \{\})\}\)
\State
\Procedure{Enumerate}{}
    \For{$i \gets 0$ \textbf{to} $n/2$}
        \State $Output \gets \{\}$
        \ForAll{$current \in Input$}
            \For{$k \gets 0$ \textbf{to} $n-1$}
                \If{$current.elements[k] = 0$}
                    \State $sum \gets current.sum + S[k]$
                    \State $expanded \gets \texttt{copy}(current)$
                    \State $expanded.elements[k] \gets 1$
                    \If{$MemoizedSums.lookup(sum) = 0$}
                        \State $MemoizedSums[sum] \gets 1$
                        \State $CanonicalSubsets[sum] \gets expanded$
                        \State $Output.add(expanded)$
                    \ElsIf{$expanded.elements < CanonicalSubsets[sum].elements$} \Comment Lex order
                        \State $CanonicalSubsets[sum] \gets expanded$
                    \EndIf
                \EndIf
            \EndFor
        \EndFor
        \State $Input \gets Output$
    \EndFor
\EndProcedure

\end{scriptsize}
\end{algorithmic}
\end{algorithm}

\begin{algorithm}[H]
\label{algo:double-meet-in-the-middle}
\caption{Subset Sum Solver Double Meet-in-the-Middle with Canonical Handling}
\adjustbox{max height=\textheight}{%
\begin{minipage}{\textwidth}
\begin{algorithmic}[1]
\begin{scriptsize}

\State \textbf{struct} \(\mathtt{kPerm(sum, elements, split, doNotExtend)}\) \{ 
  \State \quad \(elements[n] \gets \{0,0,\ldots\}\) \Comment indicator vector for inclusion
  \State \quad \(split \gets -1\) \Comment split membership (0 for \(\ell_0\), 1 for \(\ell_1\))
  \State \quad \(sum \gets 0\) \Comment sum of elements
  \State \quad \(doNotExtend \gets false\) \Comment flag for suppressing further expansion
\State \}

\State \(\ell_0 \gets \{\}\) \Comment left split
\State \(\ell_1 \gets \{\}\) \Comment right split
\State \(MemoizedSumsL0 \gets \{\}\) \Comment Unique sums from \(\ell_0\)
\State \(MemoizedSumsL1 \gets \{\}\) \Comment Unique sums from \(\ell_1\)
\State \(allKSubsetsL0 \gets \{\}\) \Comment \(k\)-subsets from \(\ell_0\)
\State \(allKSubsetsL1 \gets \{\}\) \Comment \(k\)-subsets from \(\ell_1\)
\State \(SumL0 \gets 0\) \Comment Total sum of elements in \(\ell_0\)
\State \(SumL1 \gets 0\) \Comment Total sum of elements in \(\ell_1\)

\Procedure{Initialize}{$S$}
    \For{$k \gets 0$ \textbf{to} $n-1$}
        \If{$k \bmod 2 = 0$}
            \State $\ell_0.\text{add}(S[k])$
            \State $SumL0 \gets SumL0 + S[k]$
        \Else
            \State $\ell_1.\text{add}(S[k])$
            \State $SumL1 \gets SumL1 + S[k]$
        \EndIf
    \EndFor
    \State $Input.\text{Add}(\mathtt{kPerm}(0,\{\},0,false))$ \Comment empty $k$-permutation for left split
    \State $Input.\text{Add}(\mathtt{kPerm}(0,\{\},1,false))$ \Comment empty $k$-permutation for right split
\EndProcedure

\Procedure{Solver}{$S, target$}
    \State Initialize($S$)
    \For{$i \gets 0$ \textbf{to} $\frac{n}{4} - 1$}
        \State $Output \gets \{\}$
        \ForAll{$current \in Input$}
            \If{$current.doNotExtend = true$}
                \State \textbf{continue}
            \EndIf
            \State $splitElements \gets \ell_0$ \textbf{if} $current.split = 0$, $\ell_1$ otherwise
            \State $MemoizedSums \gets \begin{cases} MemoizedSumsL0, & \text{if } current.split = 0 \\ MemoizedSumsL1, & \text{if } current.split = 1 \end{cases}$
            \State $allKSubsets \gets \begin{cases} allKSubsetsL0, & \text{if } current.split = 0 \\ allKSubsetsL1, & \text{if } current.split = 1 \end{cases}$
            \State $theOtherKSubsets \gets \begin{cases} allKSubsetsL1, & \text{if } current.split = 0 \\ allKSubsetsL0, & \text{if } current.split = 1 \end{cases}$
            \State $theOtherMemoizedSums \gets \begin{cases} MemoizedSumsL1, & \text{if } current.split = 0 \\ MemoizedSumsL0, & \text{if } current.split = 1 \end{cases}$
            \State $currentSplitSum \gets \begin{cases} SumL0, & \text{if } current.split = 0 \\ SumL1, & \text{if } current.split = 1 \end{cases}$
            \For{$k \gets 0$ \textbf{to} $\text{len}(splitElements)-1$}
                \If{$current.elements[k] = 0$}
                    \State $sum \gets current.sum + splitElements[k]$
                    \State $expanded \gets \mathtt{new}\; kPerm(sum, current.elements, current.split, false)$
                    \State $expanded.elements[k] \gets 1$
                    \If{$!MemoizedSums.contains(sum)$}
                        \State $MemoizedSums[sum] \gets expanded.elements$
                        \State $allKSubsets[sum] \gets expanded$
                        \State $Output.\text{add}(expanded)$
                        \State $\Call{Check}{expanded, target, theOtherKSubsets, theOtherMemoizedSums, currentSplitSum}$
                    \ElsIf{$\Call{IsCanonical}{expanded.elements, MemoizedSums[sum]}$}
                        \State $MemoizedSums[sum] \gets expanded.elements$
                        \State $allKSubsets[sum].doNotExtend \gets true$ \Comment suppress old
                        \State $allKSubsets[sum] \gets expanded$
                        \State $Output.\text{add}(expanded)$
                    \EndIf
                \EndIf
            \EndFor
        \EndFor
        \State $Input \gets Output$
    \EndFor
    \State \textbf{print} No solution found'
\EndProcedure

\end{scriptsize}
\end{algorithmic}
\end{minipage}%
}
\end{algorithm}

\vspace{1em}
\noindent
\textbf{Note:} This version enforces canonical subset representations using lexicographic ordering on bitmasks. This guarantees exactly one subset is retained per unique sum, ensuring correctness in high-density instances.

\begin{algorithm}[H]
\caption{Check Procedure (Canonical-Aware)}
\begin{algorithmic}[1]
\begin{scriptsize}

\Procedure{Check}{$kSubset, target, theOtherKSubsets, theOtherMemoizedSums, currentSplitSum$}    
    \If{$kSubset.sum = target$}
        \State \textbf{print} ``Subset from a single split: '' + $kSubset$
        \State \textbf{STOP}
    \EndIf

    \If{$(currentSplitSum - kSubset.sum) = target$}
        \State \textbf{print} ``Subset found on a single split using a complement subset: '' + $kSubset$
        \State \textbf{STOP}
    \EndIf

    \If{$theOtherMemoizedSums[\text{target} - kSubset.sum] = 1$}
        \State $theOther \gets theOtherKSubsets[\text{target} - kSubset.sum]$
        \If{$theOther.sum = \text{target} - kSubset.sum$}
            \State $composedSubset \gets kSubset.Combine(theOther)$
            \State \textbf{print} ``(A+B) subset sum found using both splits: '' + $composedSubset$
        \Else
            \State $composedSubset \gets kSubset.Combine(theOther.Complement())$
            \State \textbf{print} ``(A+B') subset sum found using both splits: '' + $composedSubset$
        \EndIf
        \State \textbf{STOP}
    \EndIf

    \If{$theOtherMemoizedSums[\text{target} - (currentSplitSum - kSubset.sum)] = 1$}
        \State $composedSubset \gets kSubset.Combine(theOtherKSubsets[\text{target} - (currentSplitSum - kSubset.sum)])$
        \State \textbf{print} ``(A'+B') subset sum found using both splits: '' + $composedSubset.Complement()$
        \State \textbf{STOP}
    \EndIf
\EndProcedure

\end{scriptsize}
\end{algorithmic}
\end{algorithm}

\begin{algorithm}[H]
\caption{Divide-and-Conquer Procedure (Anytime)}
\begin{algorithmic}[1]
\begin{scriptsize}
\State $ \text{ } $
\State [...]
\State // The following queues must be processed in order until depleted
\State $ schedulingQueues \gets \{\{Perm(0,\{\},0),\, KPerm(0,\{\},1)\},\, \{\},\, ...,\, \{\}\} $
\State $ lookAhead \gets n/16 $ \Comment{Empirically determined}
\State [...]
\State $ \text{ } $
\Procedure{ExpandOrSchedule}{$kPerm,\, splitElements,\, splitMemoizedSums,\, currentCycle$}	
    \State $ mustExpand \gets True $
    \State $ kIndex \gets kPerm.LastDeferralIndex + 1 $
    \State $ currentStep \gets 0 $
    \While{ $currentStep < lookAhead$ \textbf{AND} $mustExpand = True$ }
        \State $ kSubsetSum \gets kPerm.sum + splitElements[kIndex] $
        \If{ $kIndex < \text{len}(splitElements)$ \textbf{AND} $kPerm.elements[kIndex] = 0$ } \Comment{Element not yet included}
        	\If{ $splitMemoizedSums.lookup(kSubsetSum) = 1$ } \Comment{Sum already found?}
                \State $ schedulingQueues[currentCycle + currentStep].Add(kPerm) $ \Comment{Reschedule}
        		\State $ kPerm.LastDeferralIndex \gets kIndex $ \Comment{Update continuation index}
                \State $ mustExpand \gets False $ \Comment{Drop from current enumeration}
        	\EndIf
        \EndIf
        \State $ kIndex \gets kIndex + 1 $ \Comment{Next element in look-ahead block}
        \State $ currentStep \gets currentStep + 1 $ \Comment{Increment step count}
    \EndWhile
    \State \Return $ mustExpand $
\EndProcedure
\State $ \text{ } $
\State // Changes required in the Solver procedure:

\Procedure{Solver}{$S,\, target$}
	\State Initialize(\(S\))
    \State $ currentIteration \gets 0 $
	\While { $len(Input) > 0$ } \Comment{Enable multiple iterations}
		\If { $currentIteration > 0$ } \Comment{After the first iteration}
			\State $ Input \gets futureIterations[currentIteration-1] $ \Comment{Use the schedule array}
	    \EndIf
		\For { $i \gets 0$ \textbf{to} $(n/4)-1$ } \Comment{For the required columns in the split}
			\State $ Output \gets \{\} $
			\ForAll { $current \in Input$ }
    				\State [...] 
    				\State $ expand \gets true $
    				\State $ expand \gets ExpandOrSchedule(current,\, splitElements,\, currentIteration,\, MemoizedSums) $
    				\If { $expand = true$ }
                        \State [...] \Comment{Proceed with normal expansion}
    				\EndIf
			\EndFor
		\EndFor
        \State $ currentIteration \gets currentIteration+1 $
	\EndWhile
\EndProcedure
\end{scriptsize}
\end{algorithmic}
\end{algorithm}

\subsection*{Code Availability}
The source code implementing the algorithms described in this paper will be available at \\ \\ \url{https://github.com/jesus-p-salas/subset-sum-solver}.

\end{document}